# Book Chapter:
# IO Pad Integrity in Energy-Efficient Neuromorphic Chips

Arfan Ghani, Department of Computer Science and Engineering, School of Engineering and Computing, American University of Ras al Khaimah, Ras al Khaimah, United Arab Emirates

E-mail: Arfan.ghani@aurak.ac.ae

ORCID ID: https://orcid.org/0000-0002-1362-3720

## 1. Introduction

In this chapter, we take a fresh look at IOPAD integrity, highlighting its vital role in enabling low-power, high-reliability neuromorphic computing. We aim to connect the dots between the development of neuromorphic algorithms and their practical implementation in VLSI technology. This chapter dives deep into the integrity of input/output pads within the realm of low-power neuromorphic VLSI design. While much of the focus in both academia and industry has been on spiking neural networks (SNNs) and ultra-low-power core logic design, the electrical and functional integrity of the I/O interface remains a somewhat overlooked challenge. This aspect is crucial for building hardware systems that consume less energy. Throughout this chapter, we'll explore the design of the IOPAD, which is a key component of neuromorphic hardware that directly influences both signal fidelity and overall power consumption. The chapter is divided into two main sections. In the first part, we'll introduce the core concepts of SNNs and their relevance to neuromorphic computing. This is followed by the basic digital hardware design flow, from behavioural description to physical layout, for the second half of the chapter. The physical implementation phase is then discussed, emphasising the Sky Water 130 nm CMOS process as a cost-effective and industrially viable platform for neuromorphic prototyping and investigation. The chapter's attention is given to IOPAD integrity and reliability, managing the interface between the inner logic and the external world. The compromises based on power and performance aspects associated with IOPAD library selections, pad ring design approaches, and bonding configurations will be discussed. Real-world examples will be presented relative to the underlying theory principles. These investigations are framed in the broader context of energy-efficient neuromorphic systems, highlighting how stage-one I/O planning can prevent costly re-design and yield loss in later phases of development.

## 2. An Introduction to Spiking Neural Networks

### 2.1. Biological relevance

Experimental evidence has found that biologically plausible spiking neural networks could use the exact timing of action potentials for information processing, where the possibility for information representation could lie in exact spike times [1-3]. In SNNs, it is important to note that it is crucial to know how the data is processed [4-7]. The production of small action potentials or spikes represents the communication between neurons. The action potentials are around 100mV in amplitude and last for 1-2ms. The series of action potentials that occur is called a spike train, which can either occur deterministically or randomly. Once a spike has occurred, it is not possible to fire another spike immediately. The time period during which it is not possible to fire a spike is called the 'refractory period.' The simulation for an action potential is given in Figure 1.

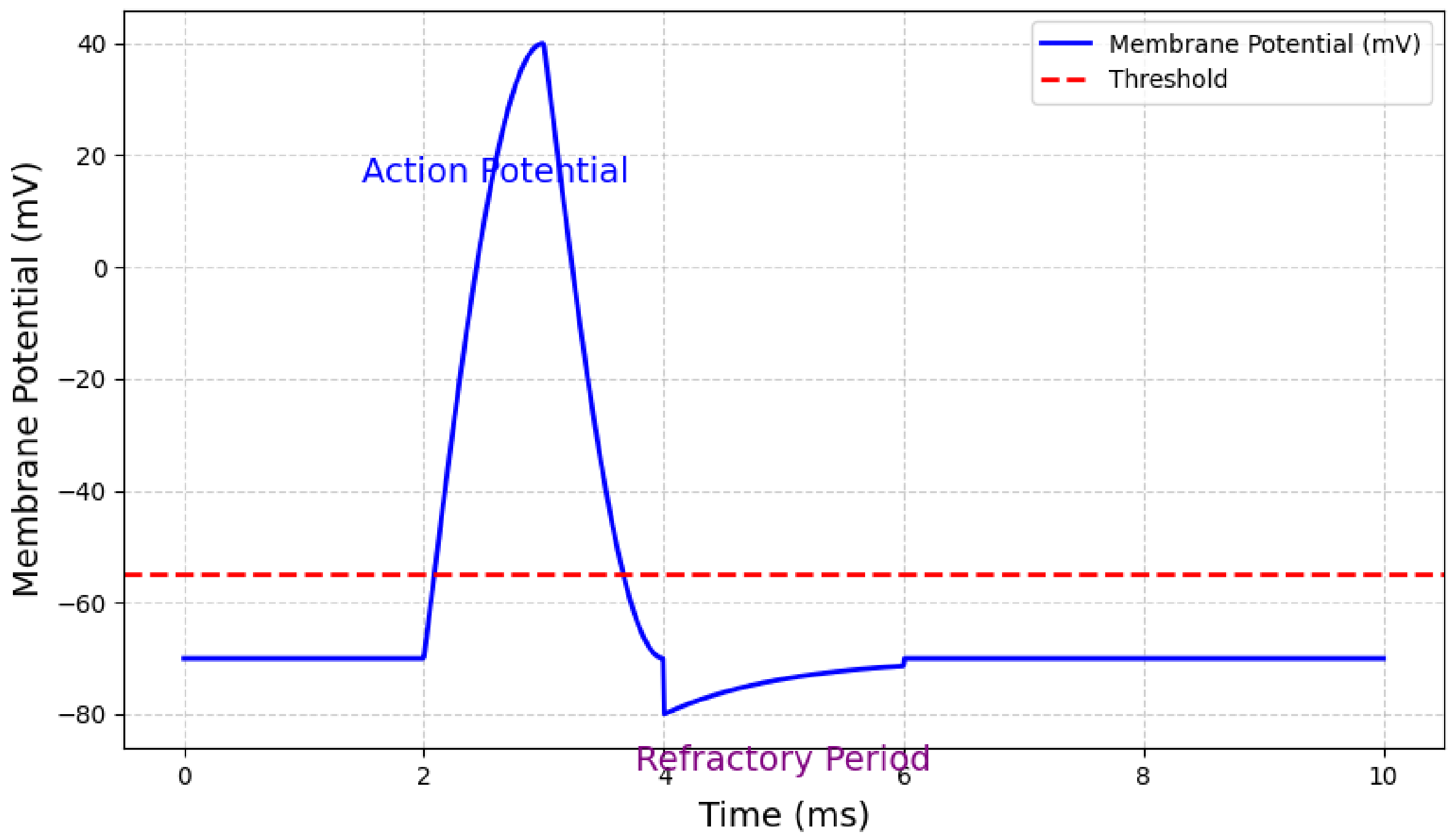

**Figure 1: An action potential [8]**

The process that happens inside a synapse starts when the action potential reaches the presynaptic terminal, initiating the release of neurotransmitter molecules. An average synapse consists of two different regions of the cell: the presynaptic membrane and postsynaptic membrane, separated by a narrow extracellular space that is called the synaptic cleft. In the presynaptic terminal, neurotransmitters are stored in synaptic vesicles, each approximately 30-40 nanometers in diameter. Such chemical messengers play a significant role in relaying signals through the synapse.

The binding of the released neurotransmitters to the postsynaptic membrane receptors can cause the opening of ion channels, leading to changes in the postsynaptic membrane potential. This can result in either the depolarisation (positive shift in membrane potential) or hyperpolarisation (negative shift) of the postsynaptic membrane, depending upon whether it is an excitatory or inhibitory signal. If the net result of the accumulation of signals exceeds a certain threshold level, then an action potential will commence in the postsynaptic neuron and then propagate through its axon to reach other neurons. The two main types of EPSPs/IPSPs, also expressed in the human brain through excitatory and inhibitory postsynaptic potentials, demonstrate the two basic types of synaptic responses, where either the postsynaptic membrane has a greater chance of action potential occurrence or a lower chance, respectively. Figure 2 illustrates EPSPs/IPSPs, depicting how different cortical neurons display complex patterns of action potentials and signal integration, investigated in greater detail by the study presented by [9].

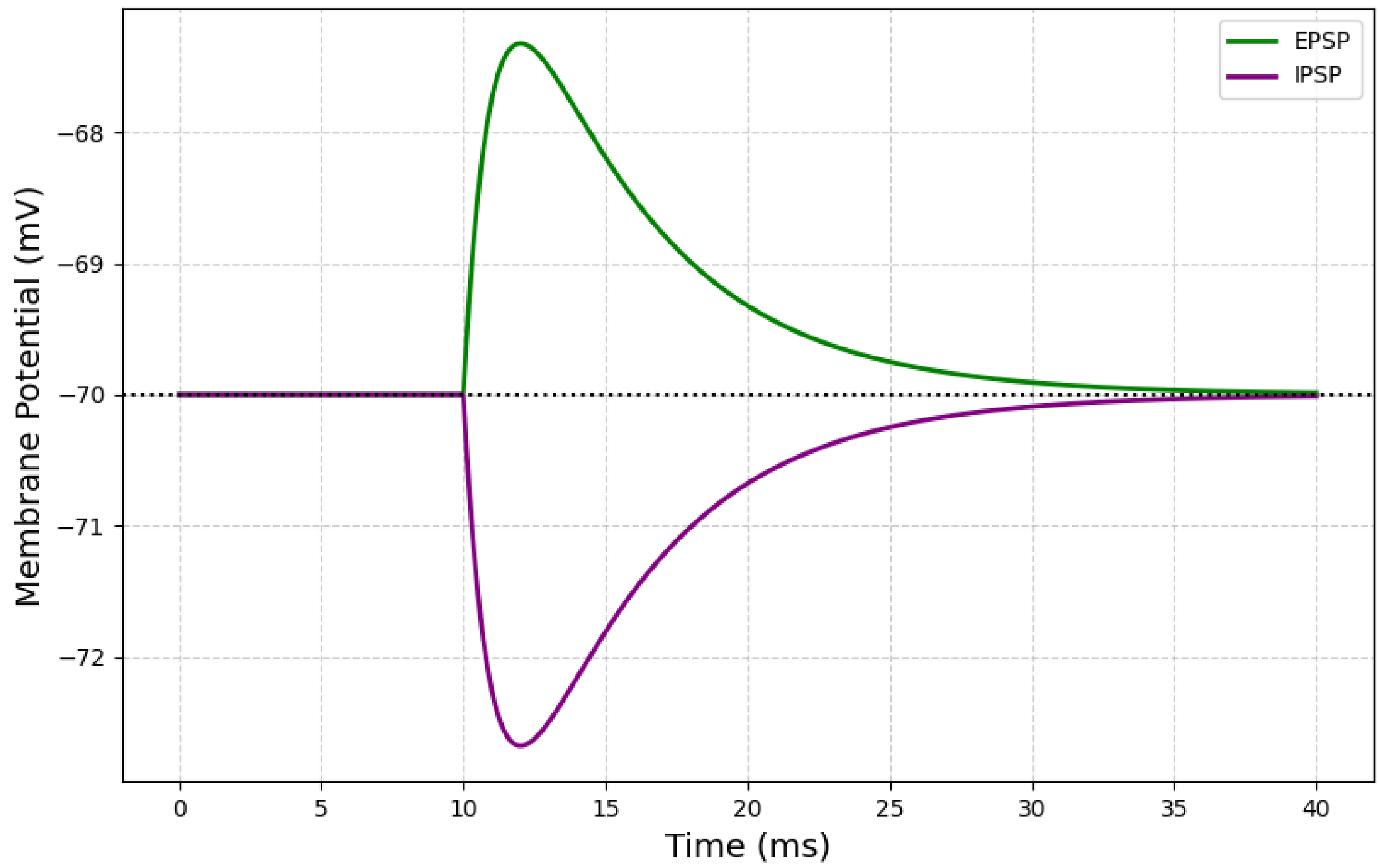

**Figure 2: Excitatory (positive) and inhibitory (negative) postsynaptic potentials**

Many types of neuron models exist in the literature, each varying from the other in terms of level of biological accuracy and complexity of simulation [10]. The type of model to choose usually depends on the level of abstraction needed for a given neural simulation task. This next section will introduce several types of spiking neuron models, along with their strengths, limitations, and complexities.

### 2.2. Hodgkin-Huxley model (HH)

One of the fundamental models in the domain of computational neuroscience is the Hodgkin-Huxley model. In a landmark study that was carried out only in 1952, scientists Hodgkin and A.F. Huxley worked on the activity of electrical currents in the squid's giant axon, providing valuable insights into the activity mechanisms in neurons [11]. This model proactively interacts through a four-dimensional system, marked by the presence of four coupled differential equations [12]. The example that describes the structural arrangement for the Hodgkin-Huxley model is presented in Figure 3.

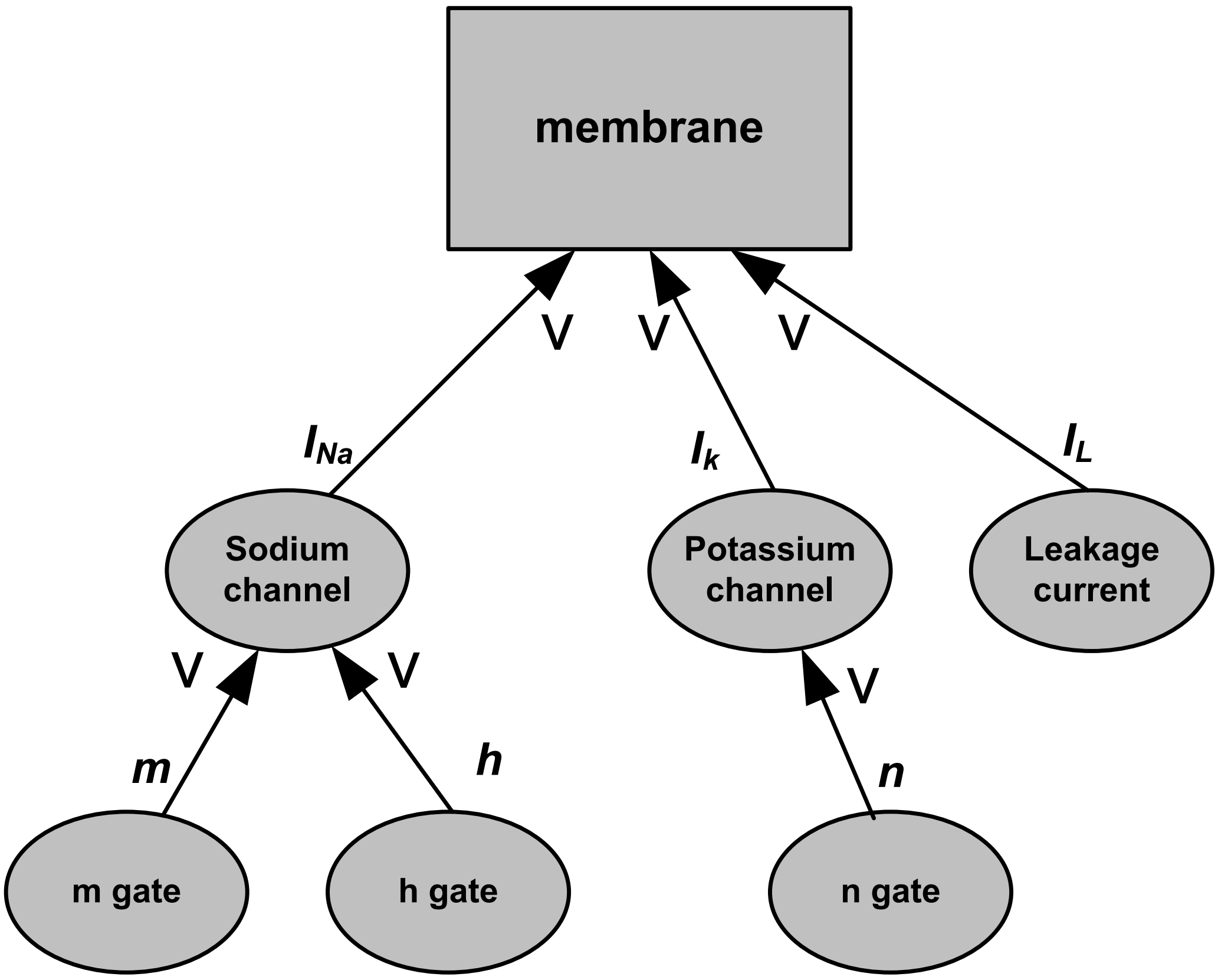

**Figure 3: Description of the HH model**

On the basis of the experimental results, the authors derived a mathematical model that could explain the electrical activity of the neuronal membrane during signal transmissions and excitation in the axon. To model the action potential in the squid nerve, Hodgkin and Huxley introduced a total of three ionic currents, namely the potassium current ($I_K$), sodium current ($I_{Na}$), and the leakage current ($I_L$), which was subsequently identified mainly with chloride ions. The capacitance of the membrane is called $C_m$, while $V_{Na}$, $V_K$, and $V_L$ are the reversal potentials for sodium, potassium, and leakage currents, respectively. The expression for the total membrane current, $I_m$, is given by equation (1) and is illustrated graphically in Fig. 4.

$$I_m = C_m \frac{dV}{dt} + I_{Na} + I_K + I_L \qquad \text{Eq-(1)}$$

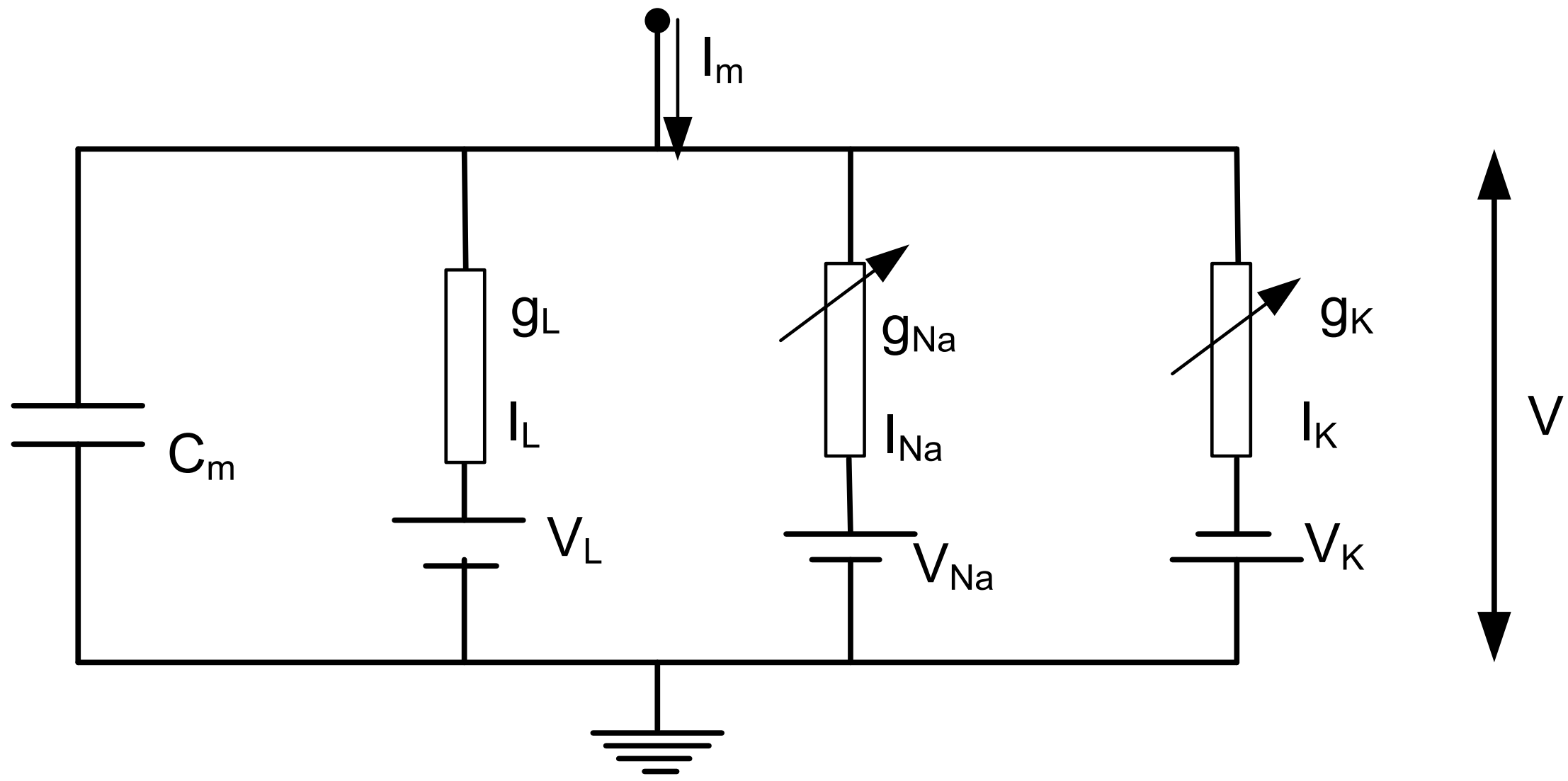

**Figure 4: HH equivalent electrical circuit**

The expressions for ionic currents are outlined below.

$I_K = G_K\, \eta^4 (U - E_K)$ Eq - (2)

$I_{Na} = G_{Na}\, \mu^3\, \theta\, (U - E_{Na})$ Eq - (3)

$I_L = G_L (U - E_L)$ Eq - (4)

Here, $G_K$, $G_{Na}$, and $G_L$ denote the maximal conductances for the potassium, sodium, and leak channels, respectively. The terms $E_K$, $E_{Na}$, and $E_L$ represent their corresponding reversal (equilibrium) potentials, while U stands for the transmembrane voltage relative to a resting reference level. Table 1 summarises the typical parameter values, adapted from the Hodgkin-Huxley framework, assuming a reference point where the resting potential is set to zero.

**Table 1: Parameters of the HH Model**

| **Channel** | $E_x$ *(mV)* | $G_x$ *(mS/cm²)* |
|---|---|---|
| **Na** | 115 | 120 |
| **K** | -12 | 36 |
| **L** | 10.6 | 0.3 |

The temporal dynamics of the gating variables are captured using first-order kinetic equations, in which the forward alpha and beta transition rates are functions of voltage and were originally fitted by Hodgkin and Huxley to match experimental squid axon data.

$$d\mu/dt = \alpha_\mu (U) (1-\mu) - \beta_\mu (U)\mu \quad \text{Eq- (5)}$$
$$d\eta/dt = \alpha_\eta (U) (1-\eta) - \beta_\eta (U)\eta \quad \text{Eq- (6)}$$
$$d\theta/dt = \alpha_\theta (U) (1-\theta) - \beta_\theta (U)\theta \quad \text{Eq- (7)}$$

The voltage-dependent rate functions ($\alpha$ and β) for each gating variable are listed in Table 2, where exponential terms and rational expressions capture the probabilistic nature of channel opening and closing.

**Table 2: Rate functions for gating variables**

| *Variables* | $\alpha_x (U)$ | $B_x (U)$ |
|---|---|---|
| $\eta$ (K) | $(0.1-0.01U)/[\exp(1-0.1U)-1]$ | $0.125\exp(-U/80)$ |
| $\mu$ (Na activation) | $(2.5-0.1U)/[\exp(2.5-0.1U)-1]$ | $4\exp(-U/18)$ |
| $\theta$ (Na inactivation) | $0.07\exp(-U/20)$ | $1/\exp(3-0.1U)+1$ |

The Matlab output, presented in Figure 5, shows that a rise in the membrane voltage 'v' causes a reduction in the 'h' variable, along with an 'm' and 'n' augmentation. This implies that when the stimulation from the outside increases the cell's membrane, 'm' also increases, thereby causing sodium ions to rush inside the cell. If such depolarization reaches a threshold level, it causes an action potential. The output also portrays that the sodium conductance will eventually saturate for higher values of 'v' because 'h' is diminishing where the time constant for 'm' is smaller than that for 'h.' Since the time constant for *m* is smaller than that of *h*, *m* responds more rapidly to changes in voltage, while *h* lags. Additionally, potassium current begins to flow outward on a comparable time scale. Figure 6 presents the spike initiation and its numerical integration using Euler's method with a time step of 0.01.

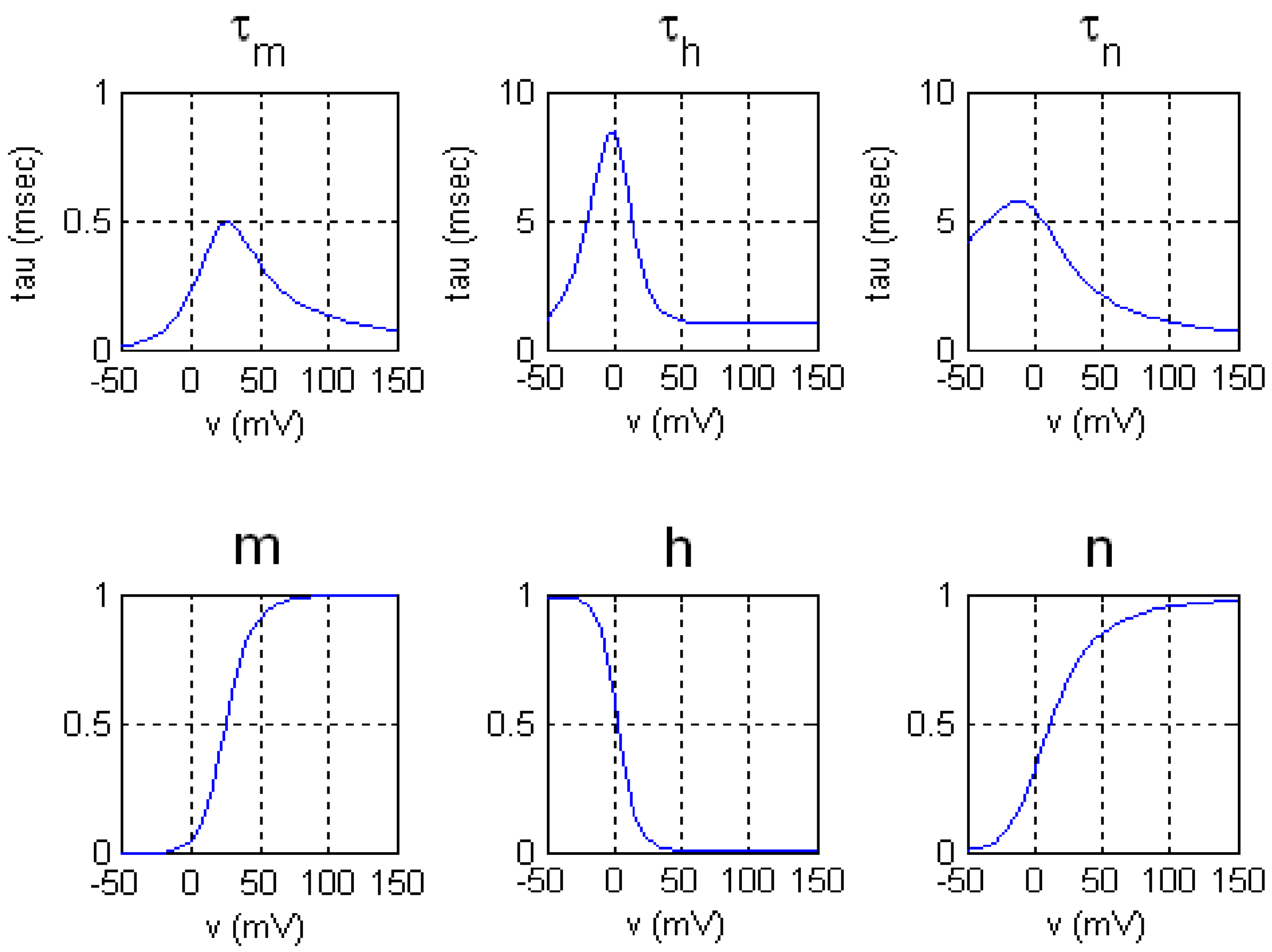

**Figure 5: HH model and the impact of time constant.**

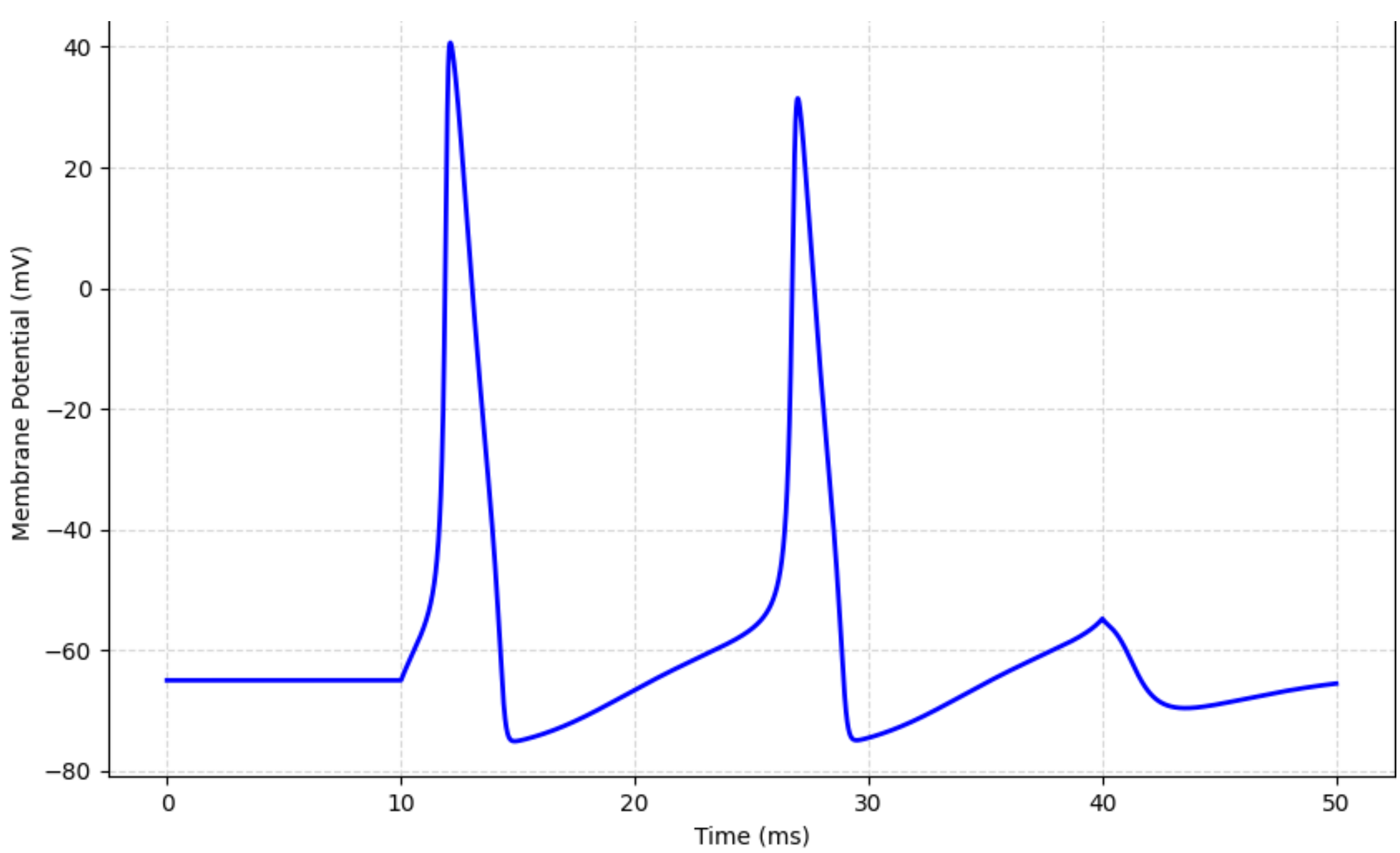

**Figure 6: Integration of the HH model with Euler integration**

Models based on conductance are well-suited for describing neural phenomena such as action potentials, spike generation, thresholds, bursting behaviour, synaptic effects, and spike adaptation. However, a major drawback of these models is their complexity, due to the high number of parameters involved. They are also computationally demanding and perform poorly when run on traditional sequential computing systems. In the sections that follow, several simplified or abstract neuron models are discussed. These alternatives are more computationally efficient than Hodgkin-Huxley-type models, while still capturing many essential features of real neurons.

### 2.3. Integrate and fire model (I&F)

Figure 7 depicts the fundamental structure for modelling the Integrate-and-Fire (I&F) model. The capacitor in the model integrates the arriving current I(t). Then, when it reaches a threshold value θ, the neuron will fire a spike, and the capacitor voltage level is instantly reset to zero. The mathematical expression for the I&F model is given below.

$$I(t) = C_m \frac{du_i(t)}{dt} \qquad \text{Eq-(8)}$$

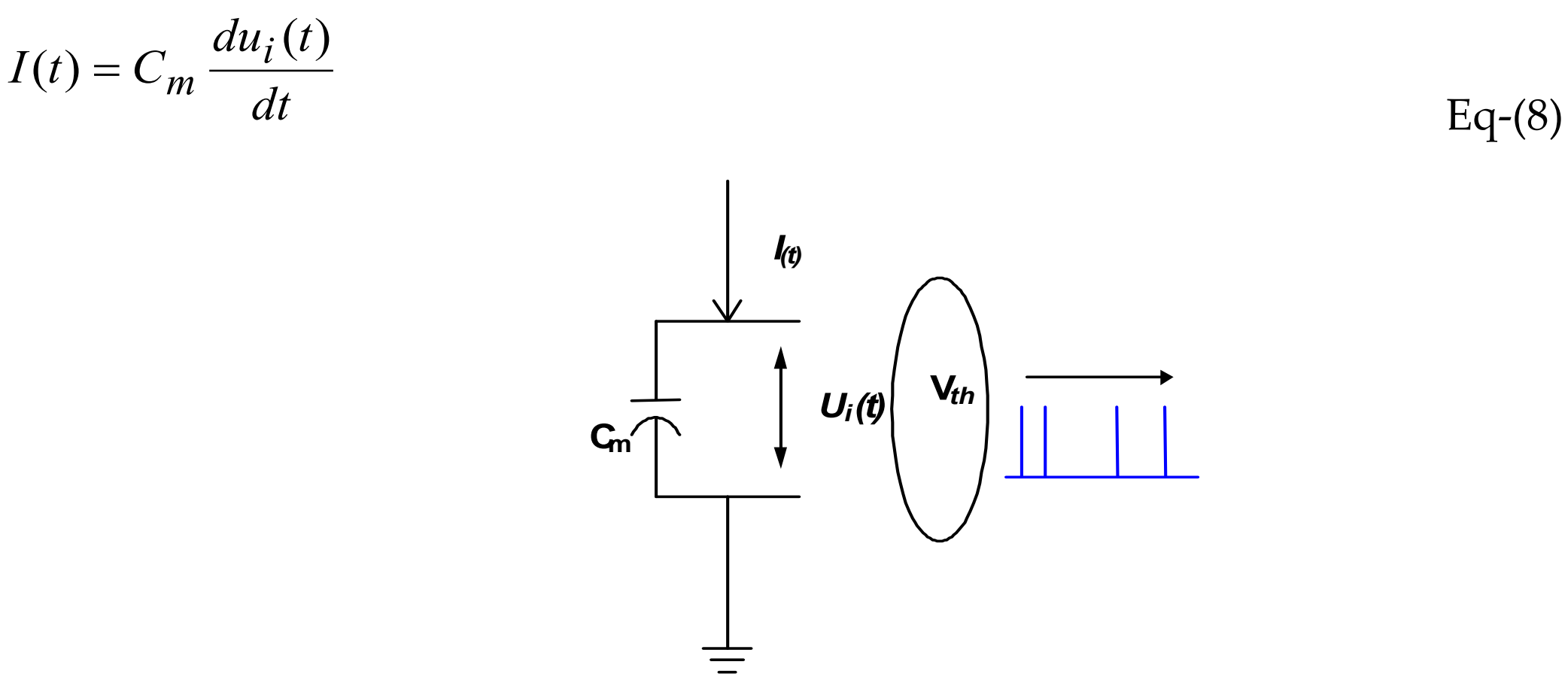

**Figure 7: Integrate-and-fire model without leakage**

In this mathematical model, 'Cm' is the capacitance of the membrane, '*Ui(t)*' is the membrane potential, and '*I(t)*' is the input current. The model describes how the neuron integrates stimuli and fires when the membrane potential exceeds a certain threshold level. However, despite its simplicity, it is an unrealistic model from a biological perspective, since actual biological neurons cannot integrate stimuli in zero time. Second, the I&F model does not account for the existence of leakage current, which is a critical parameter when describing the time-resolved behaviour of the membrane's charging and discharging process. The activity pattern of an I&F neuron is presented in Fig. 8.

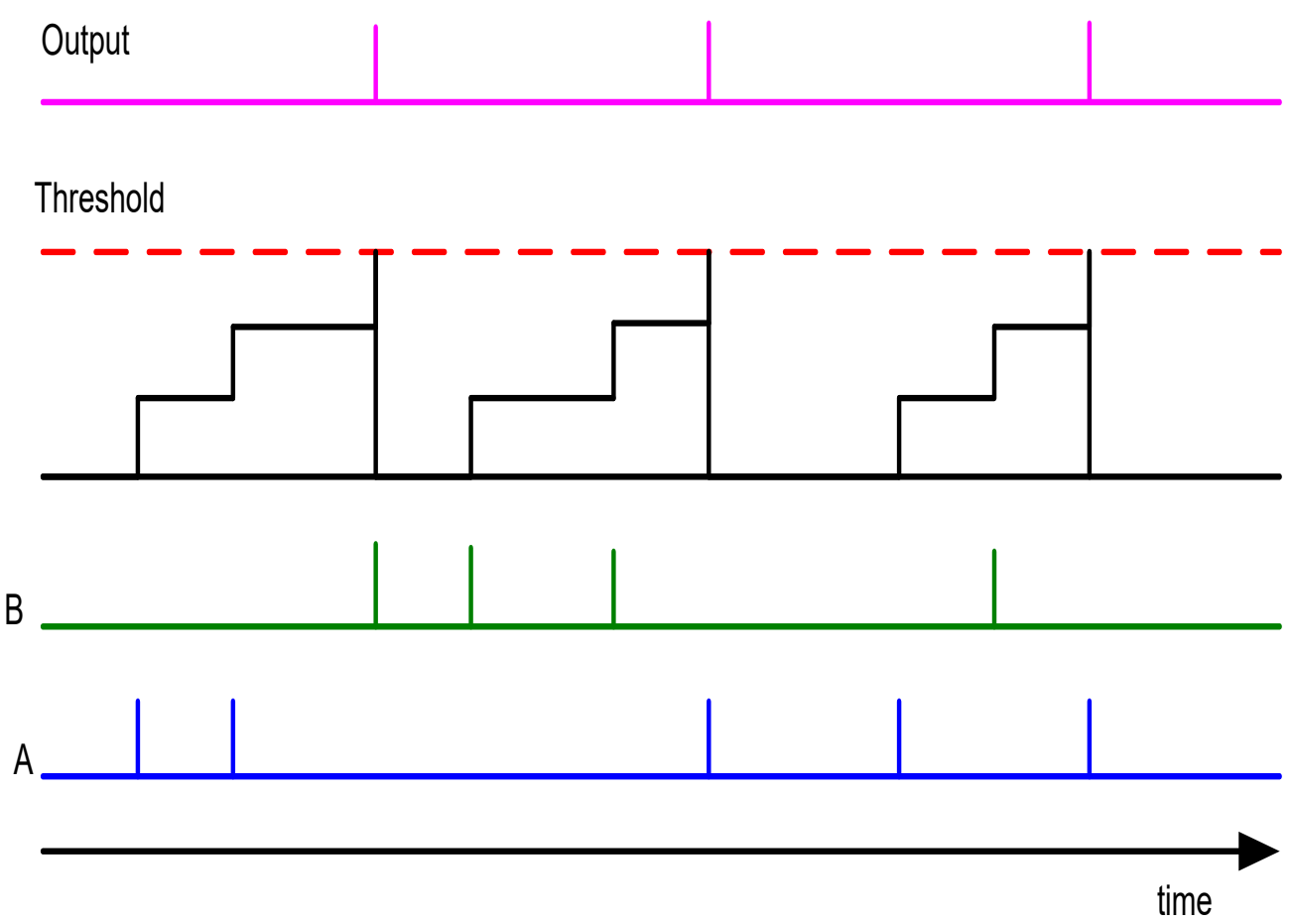

**Figure 8: Behaviour of an integrate and fire neuron.**

## 2.4. Integrate and fire model (LIF) with leakage

The leaky integrate-and-fire (LIF) model is commonly used to describe the firing behaviour of neurons based on key characteristics such as the membrane time constant, firing threshold, and refractory period [13]. One of the major simplifications in this model is that it ignores the detailed shape of the action potential, focusing solely on the precise timing of spikes. In contrast to idealised integrate-and-fire models, biologically plausible neurons exhibit a behaviour of leaky integrator that gradually brings the membrane potential back to its resting state when no input is present. This leakage effect is modelled by including a membrane resistance *Rm*, as illustrated in Figure 9.

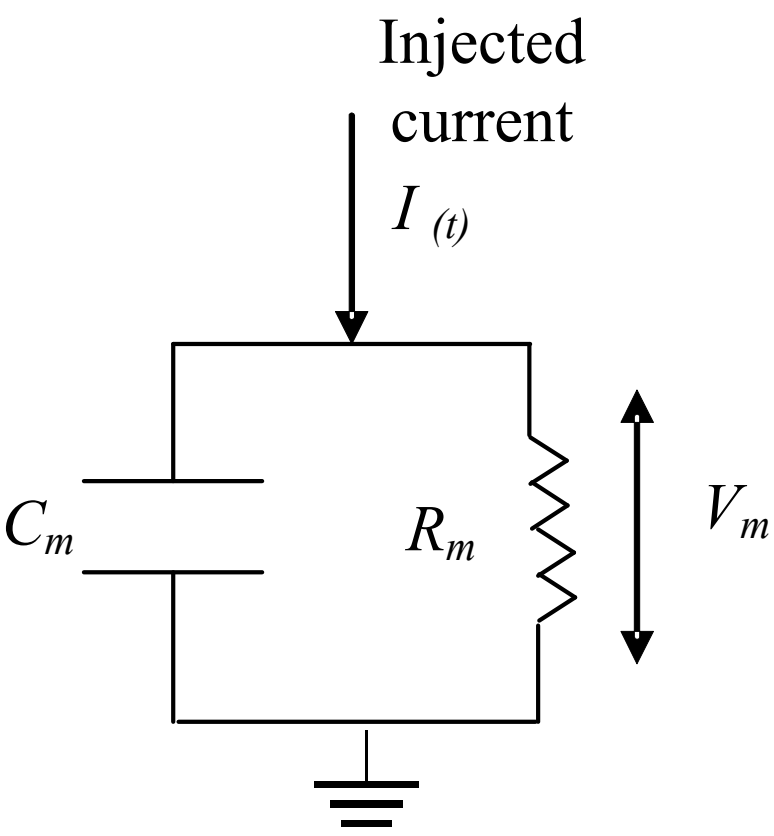

**Figure 9: An integrate-and-fire neuron model with leakage**

A mathematical representation of this model is shown as under:

$$I(t) = C_m \cdot \frac{dV_m}{dt} + \frac{V_m}{R_m} \qquad \text{Eq- (9)}$$

Figure 10 shows the MATLAB simulation results for a leaky integrate-and-fire (LIF) neuron. In this model, the incoming signals are integrated over time, and when the membrane potential surpasses a predefined threshold *Vth*, it generates a spike and immediately resets to a resting level *Vreset*. Although LIF neurons can replicate a few core neurocomputational features—specifically synaptic integration, reset behaviour, and membrane threshold - their capabilities are limited compared to more detailed models. As noted by [14], the LIF model accounts for only 3 of the 20 key properties observed in biological neurons. Being a one-dimensional representation, it lacks many essential characteristics of cortical spiking behaviour. Nonetheless, it remains useful for simulating large-scale neocortical activity in cases where exact biological fidelity is not required.

Because simplified neuron models enhance conceptual clarity and ease of analysis [15-16], the LIF model has been widely adopted in various theoretical and computational studies [12-13], [17 - 20]. Its straightforward implementation has also made it a popular choice in a range of engineering applications [21 - 24].

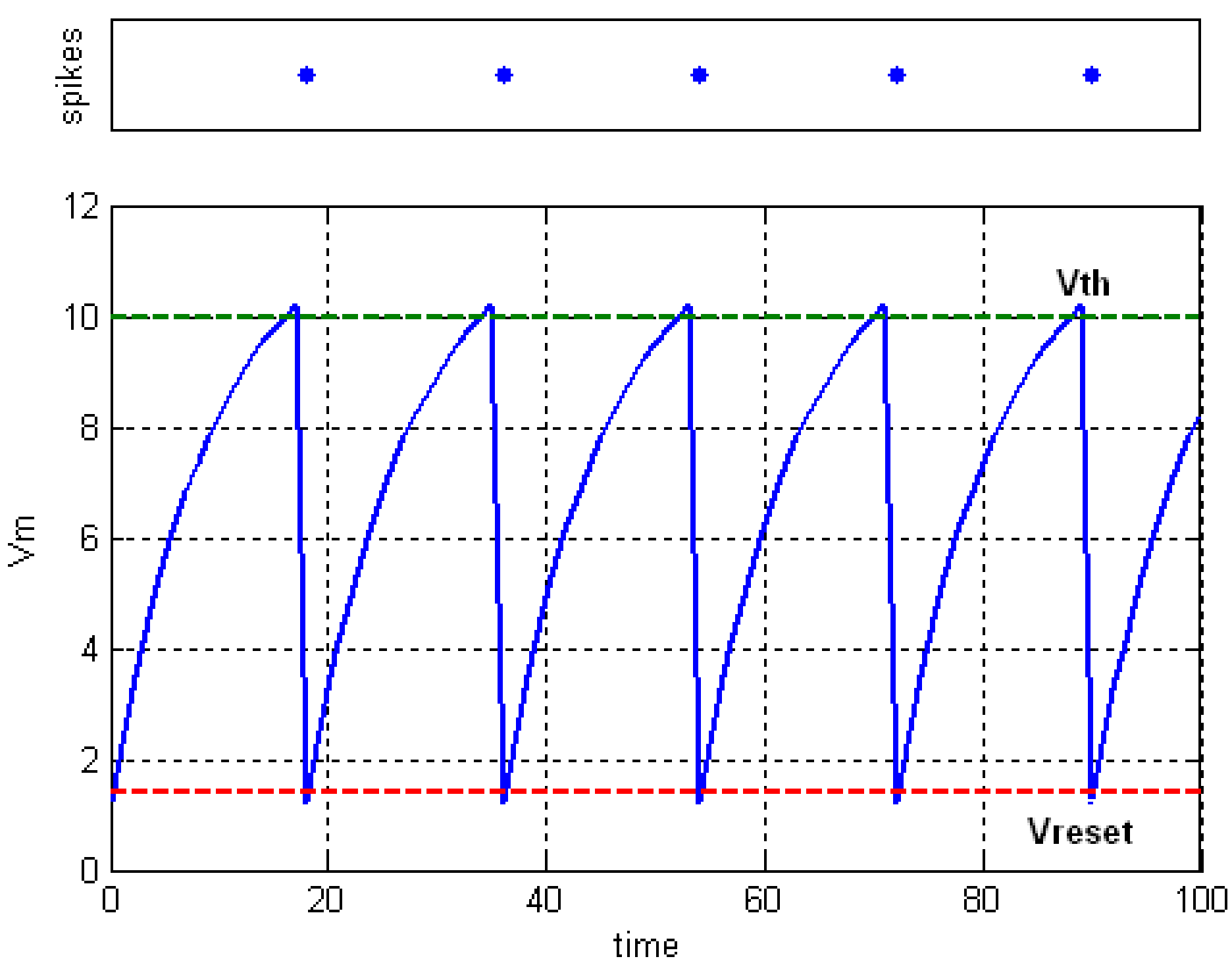

**Figure 10: Matlab simulations of the LIF neuron**

## 2.5. Spike response model (SRM)

The spike response model (SRM) represents an extension of the leaky integrate-and-fire (LIF) model, offering more flexibility in its modeling process concerning the behavior of neurons. The SRM also differs from LIF models, non-linear or otherwise, by using the time passed since the last spike instead of functions related to the membrane potential. The other major difference is in the form used to mathematically model the elements, wherein integrate-and-fire models require the use of differential equations, whereas SRM equations include the membrane potential at time *t* by an integral that relates to past activity [12].

The mathematical representation of the SRM is given below.

$$u_i(t) = \sum_{t_i^{(f)} \in F_i} \eta_i(t - t_i^{(f)}) + \sum_{j \in \Gamma_i} \sum_{t_j^{(f)} \in F_j} w_{ij} \varepsilon_{ij}(t - t_i^{(f)}) \quad \text{Eq-(10)}$$

The symbol $\eta i(s)$ models the refractory behaviour of the neuron. It is used to reset the neuron and gradually decays to zero over time. For any time $s < 0$, $\eta i\ (s) = 0$. The $\varepsilon ij$ (s) is called the form of the post-synaptic potential (PSP). This depends on the spike activity of the presynaptic neurons and affects the membrane potential of neuron *i*. This is zero for $s < 0$. The state of any neuron at time *t* is denoted by $ui\ (t)$. The strength, or weights, $wij$, modulate the amplitude of the PSP. The collection of firing times for neuron *j* is denoted by *Fj*, and it is defined for the set of pre-synaptic neurons to neuron *i*, denoted by *i*.

*Ui(t)* does not change from its resting level when there are no incoming spikes, according to the model. The membrane potential is altered by a contribution denoted by $\varepsilon ij$ for each incoming spike, modeling how the neuron responds to an incoming spike. The membrane potential increases beyond a threshold denoted by $\theta$, causing the neuron to generate an output spike, after which the membrane potential resets to its resting level. To model the refractoriness phenomenon, an additional parameter $\eta$ is added during this reset process, ensuring that there are no new inputs during the period.

Figure 11 below illustrates the process that takes place in the spike response model. The incoming spikes are added at different time intervals, hence contributing to the membrane potential. The dotted line in the figure signifies the threshold, beyond which, when the total potential is exceeded, an output post-synaptic potential is produced. The silent period of the neuron, depending on $\eta$, comes after the production of the spike output.

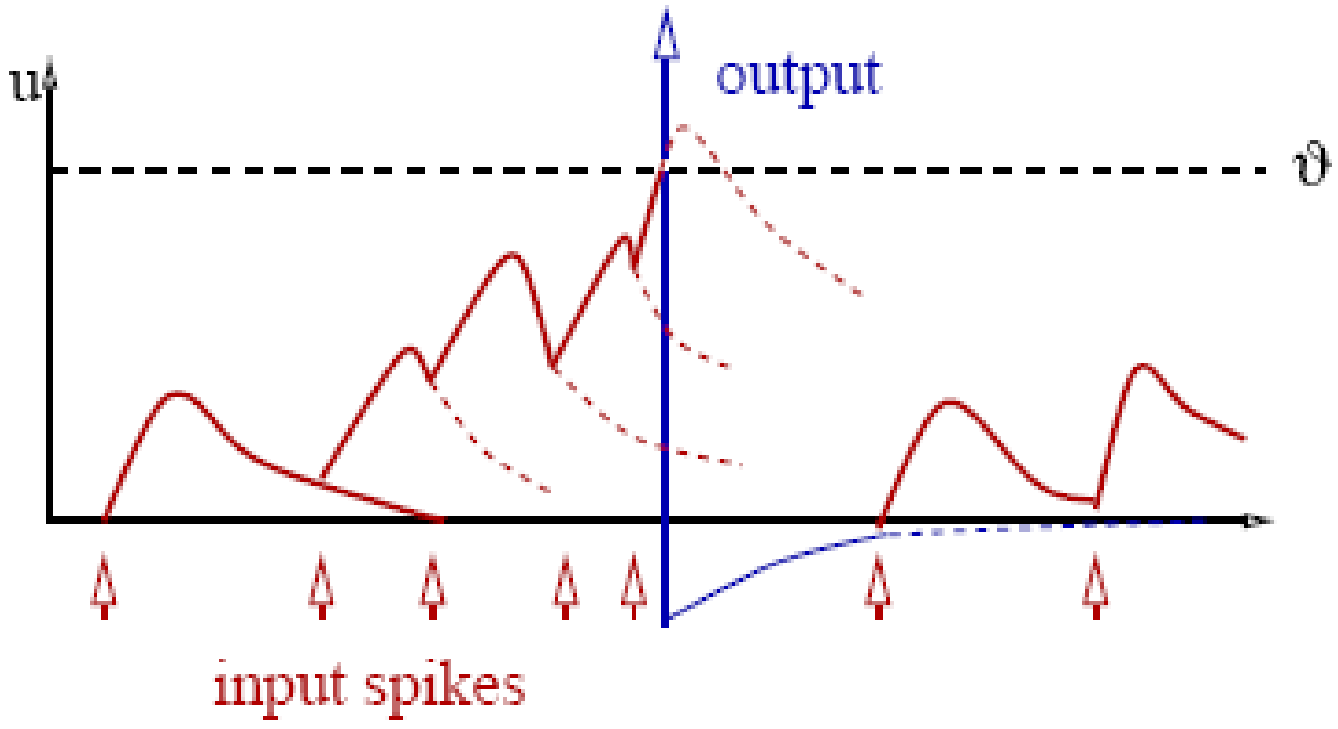

**Figure 11: Spike response model.**

### 2.6. Izhikevich spiking neuron model

The Izhikevich neuron model is a biologically realistic two-dimensional model that can be described using coupled differential equations. It is capable of reproducing the firing patterns observed in all known types of cortical neurons [25]. The dynamics of this model are governed by two interrelated differential equations:

$$\frac{du}{dt} = 0.04u^2(t) + 5u(t) + 140 - w(t) + I(t) \qquad \text{Eq. (11)}$$

$$\frac{dw}{dt} = a(bu(t) - w(t)) \qquad \text{Eq. (12)}$$

Spike resetting is represented as shown below:

$$\text{If } u \geq \theta \text{ then u} \leftarrow c \qquad \text{Eq. (13)}$$

and $w \leftarrow w+d$ Eq. (14)

This model is defined using the variables $u$, denoting the membrane potential, and $w$, denoting the recovery variable. The parameters $a$, $b$, $c$, and $d$ are also dimensionless. To guarantee that $u$ is in millivolts ($mV$) and time is in milliseconds ($ms$), the expression chosen is $0.04u^2(t) + 5u(t) + 140$. With different values for the chosen parameters, it's possible to create different patterns using the model, illustrated in Figure 12 and discussed by [25].

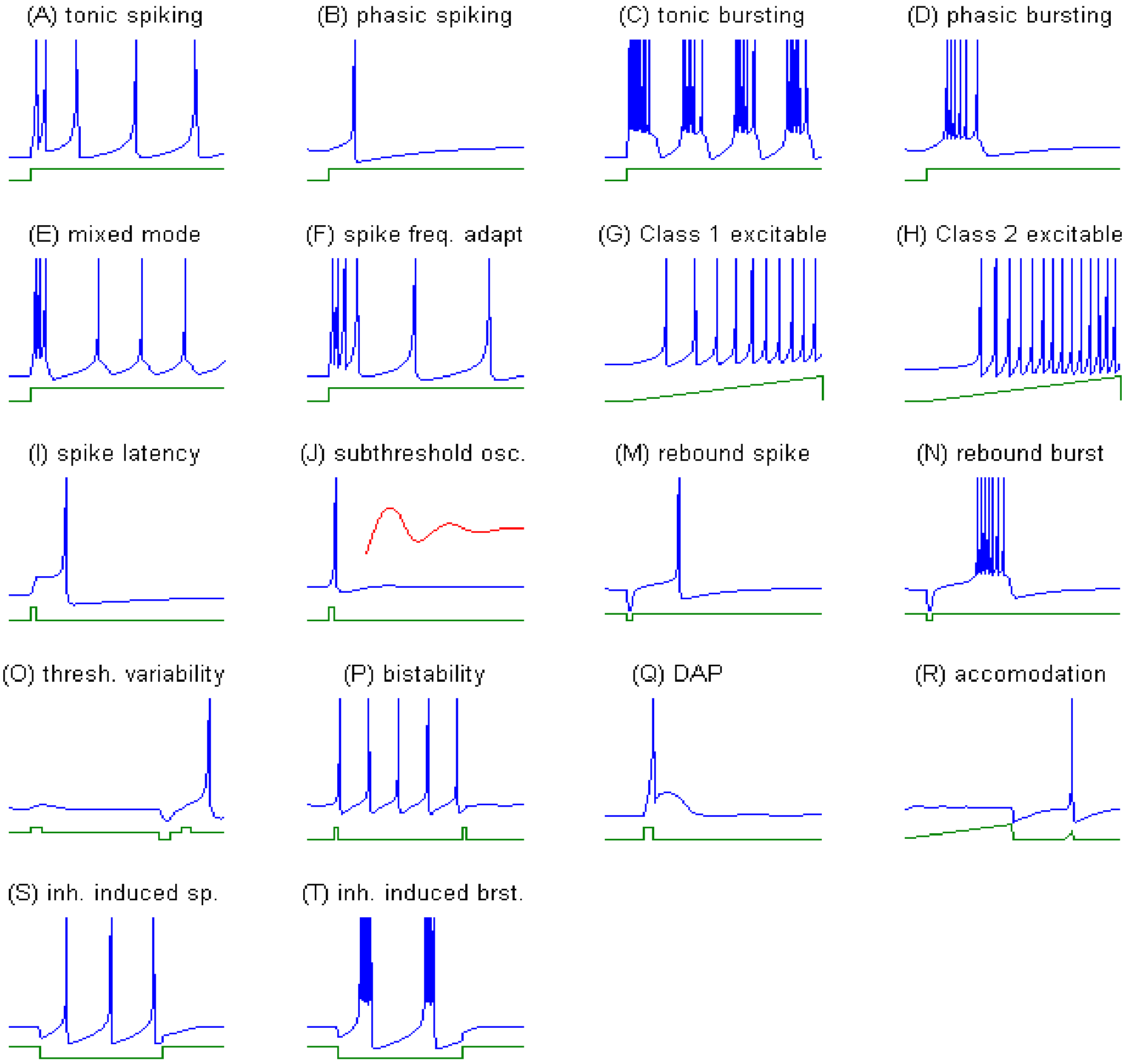

**Figure 12: Izhikevich neuron models [14].**

### 2.7. Discussion

The various types of spiking models for neurons were discussed in this section of the book chapter. The type chosen depends on how much biological accuracy is required, weighed against how computationally efficient it needs to be. The type that has a lot of biological accuracy, such as that offered by the Hodgkin-Huxley (HH) or similar models, has too much biological accuracy for something like neural computation, where it has to be fast and computationally efficient. Also, such models are too computationally expensive, requiring too much processing power to compute. Thus, it is not so useful when trying to compute artificial neural networks for use in machine learning or engineering applications on standard computer processing units, since it will require too much processing power, making it computationally expensive. Other models that are less computationally expensive, such as the Integrate-and-Fire (I&F) model, are too simple to compute and lack too much biological accuracy. They only compute the time difference for the neurons' firing, making it simple to compute since it does not compute too many biological factors present in actual neurons. This model requires low processing power, making it easier to compute artificial neural networks for use in machine learning or engineering applications. The tradeoff in model type from too much biological accuracy to too little biological accuracy, depending on whether it is for machine learning or engineering applications, may not be that straightforward. If it were for studying the biological process present behind how actual neurons function, then models such as the Hodgkin-Huxley model would be the better option since it has a greater level of biological accuracy, providing a more accurate model and representation. But for machine learning or engineering applications, artificial neural networks requiring lower processing power, such as the I&F model, will provide better results since it will only require less processing power to compute artificial neural networks for use in machine learning or engineering applications.

The true strength in neural networks, therefore, is derived from their learning capabilities. Whether it is learning by distinguishing patterns in data or by making decisions through past experiences, it is learning that enables neural networks to become so valuable. Learning algorithms also enable neural networks to become better through time, and it is important for learning algorithms in neural networks to be fully understood for applications that require neural networks to undertake more complex tasks. In the next section, we will explore how learning takes place in Spiking Neural Networks.

## 3. Learning in Spiking Neural Networks

The synaptic plasticity hypothesis, whose origins date back to the original research of Donald Hebb (1949), has long been the underlying model of learning within artificial neural networks. Hebbian plasticity, as especially defined through long-term potentiation (LTP) and depression, is the basis for the majority of models of learning. Spike Timing Dependent Plasticity (STDP) is one of the most important mechanisms in this paradigm, a Hebbian plasticity that is time-locked to the spikes in the neural activity. In STDP, if a presynaptic spike is followed by a postsynaptic spike, then the synapse gets reinforced, and if followed by a postsynaptic spike, then the synapse gets depressed. The durability of LTP, or the synaptic memory, is a function of time that decreases exponentially with time, determined by the membrane's time constant. This means that only presynaptic spikes arriving within a specific time window relative to the postsynaptic spike contribute to synaptic potentiation. Conversely, if presynaptic spikes arrive shortly after a postsynaptic action potential, the synapse experiences depression. When the timing of the presynaptic and postsynaptic spikes is too far apart, there is no change in the synaptic weight. The most commonly used timing windows for regulating synaptic weights were derived from experiments involving cultured rat hippocampal neurons, as reported by [26]. I conducted simulations to replicate the behaviour of STDP, and the results are shown in Figure 13. In the figure, the upper right section illustrates Long-Term Potentiation (LTP), while the lower left section represents Long-Term Depression (LTD). These simulations are valuable as they provide insights into the possibility of designing a circuit based on this synapse model for on-chip learning applications.

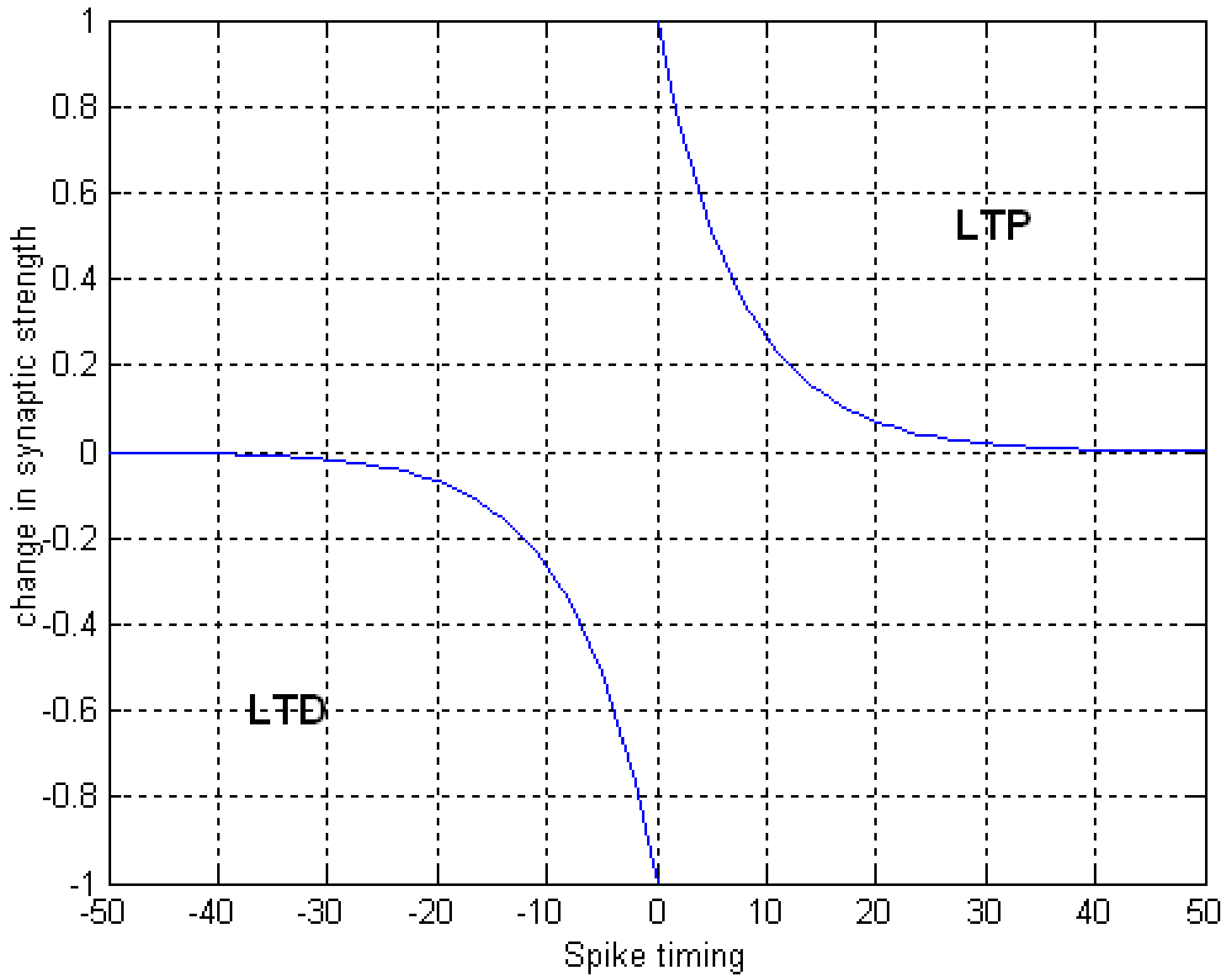

**Figure 13: Matlab simulation of an STDP-based plasticity**

STDP can be modelled by approximating the contribution of both presynaptic and postsynaptic spikes to the synaptic change. The synaptic modification over time can be expressed using exponential functions:

For $t < 0$, the change is given by $A+ \, exp(t/\tau+)$, where $A+$ is greater than 0.

For $t \geq 0$, the change follows $A- \, exp(-t/\tau-)$, with $A-$ being less than 0.

The parameters $\tau+$ and $\tau-$ define the time window during which synaptic changes can occur, depending on the timing of the presynaptic and postsynaptic spikes. The constants $A+$ and $A-$ set the maximum possible change in synaptic strength, with the most significant changes happening when the presynaptic and postsynaptic spikes occur in proximity. In Figure 13, the spike timing is represented by the difference in firing times between the presynaptic and postsynaptic neurons. When the time difference $\Delta t$ is greater than 0 and near zero, the synaptic weight increases. The synaptic change ($\Delta w$) is represented as a percentage. Various forms of STDP have been discussed in the literature, and the main differences among these models lie in the symmetry between LTP and LTD, as well as discontinuities observed in the $\Delta w$ function.

### 3.1. Hebbian Learning

One of the most well-known learning principles was proposed by the Canadian neuropsychologist Donald O. Hebb in 1949. His work introduced a concept that explains how learning occurs through the synaptic connections between neurons. Hebb's rule states: "When the axon of cell A is close enough to excite cell B and consistently participates in firing it, some growth or metabolic changes occur in one or both cells, such that the effectiveness of A in firing B is enhanced." In simpler terms, when two neurons connected by a synapse are activated within a specific time window, the strength of that synapse increases. However, if the neurons are activated outside of this time window, the synaptic strength is weakened and the learning process based on this principle is known as Hebbian learning. Hebbian learning has several key characteristics, such as being unsupervised and occurring locally, which adds to its biological plausibility. It is a concept that could be useful when benchmarking or designing circuits for both hardware and software implementations.

Mathematically, Hebb's rule for adjusting the weight $w_{ij}$ from neuron $j$ to neuron $i$, with a learning rate $\eta$, can be expressed as follows and shown in Figure 14(a):

$$\Delta w_{ij} = \eta \, x_i \, x_j$$

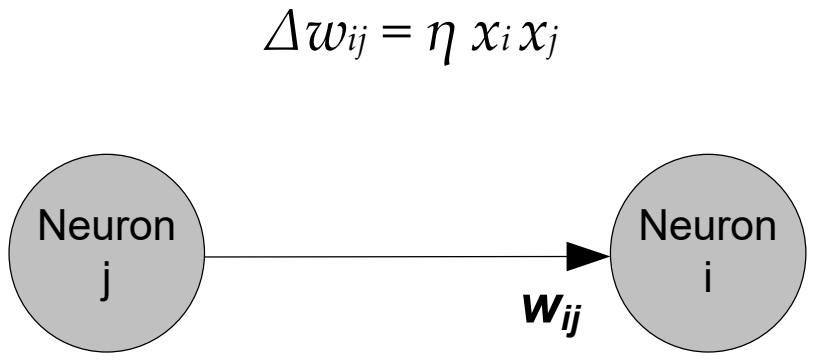

**Fig. 14(a). Hebbian Learning**

The output of the *i-th* neuron is given by $X_i = F\,[W_{ij}\, X_j]$, where $F$ is a linear or nonlinear function to transform the input into an output. The change in synaptic weight, $\Delta w_{ij}$, is given as $\eta\, x_i\, x_j$, where $\eta$ is the learning rate. The Hebbian learning rule is associative because the presynaptic and postsynaptic neurons need to be activated within a limited time period to change the synaptic weight. One of the limitations of this rule is that there is no bound on the connection weight, which results in unbounded growth. But then the issue can be rectified by including a decay function to every connection such that the weights remain within reasonable bounds.

## 4. VLSI Design Considerations for Spiking Neural Networks

The aim of this section is to discuss the realisation of a power-aware hardware platform which can imitate the characteristics of spiking neurons with very good accuracy. The circuit of neuro-inspired VLSI is based on the studies of biological neural systems and tries to imitate the capability of neural networks at the hardware level. These chips' energy consumption is minimised because they are designed to operate at low power, and they compute by using the parallelism and temporal dynamics of spiking neurons in an energy-efficient way. The design of the circuits for the neuromorphic VLSI incorporates various aspects, such as STDP, changing of synaptic weights and generating spikes dynamically. By devoting hardware to the simulation of these processes, neuromorphic VLSI chips provide a performance and energy efficiency level that is vastly superior to classic digital processors for applications in sensory processing, learning, and pattern recognition [27-28].

### 4.1. LIF Neurons and Circuit Simulations

In order to show the functionality of SNN at the circuit level, we are examining the behaviour of hardware spiking neural networks (SNNs) made with leaky integrate-and-fire (LIF) neurons using Verilog hardware description language. Our main focus will be on circuits' hardware behaviour and how they can be optimised for energy-efficient implementation. A spiking input neuron was connected to an LIF neuron through a dynamic synapse model that simulates both temporal and synaptic dynamics in a biological-like way. The setup depicted in Figure 14 (b) was executed with Verilog code, which resulted in detailed waveform simulations that helped us understand the temporal behaviour and response of the network. The simulation captures various important traces, such as the membrane voltages of the LIF neuron, the postsynaptic potential (PSP) produced at the synapse, and the output spikes that are generated by the neuron in reaction to incoming inputs. The membrane potential of the LIF neuron collects the postsynaptic response over time, and when this potential surpasses a previously set threshold, an action potential occurs. This spike, in turn, indicates the neuron's firing behaviour, which is very important for neural network calculations.

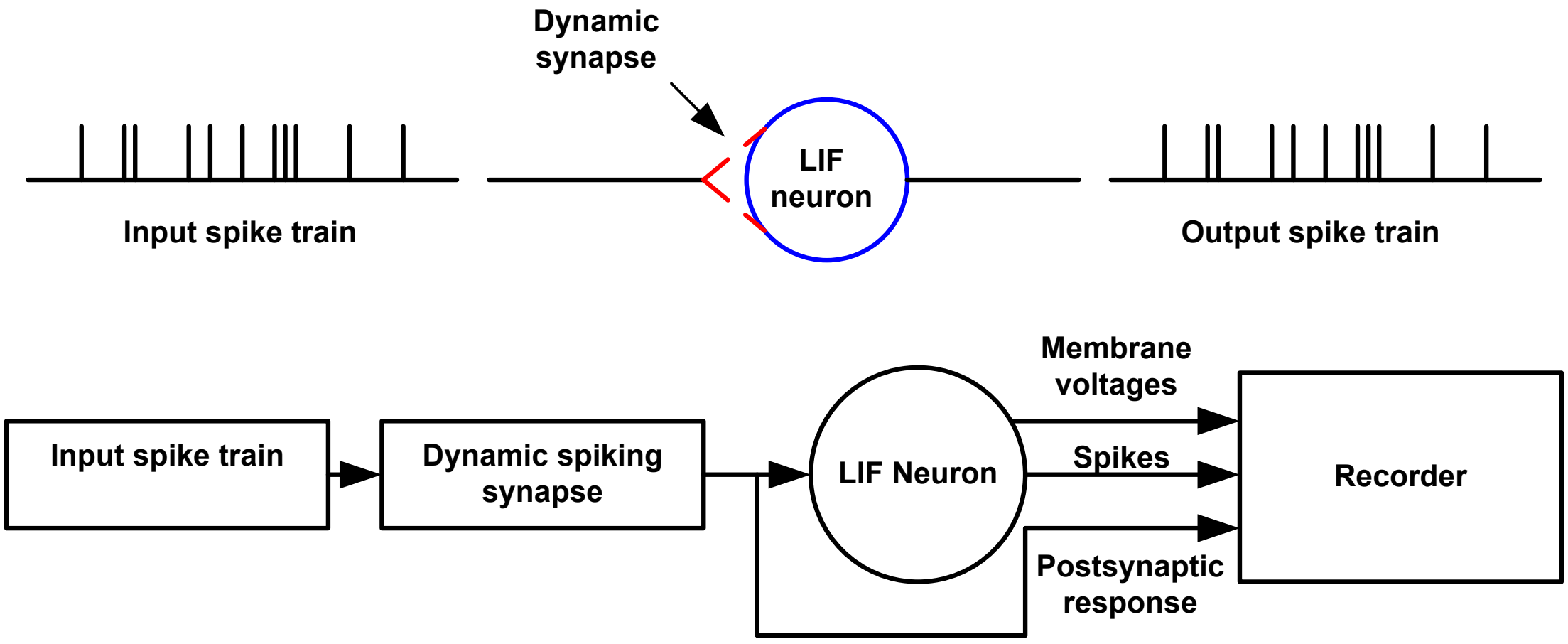

**Figure 14 (b): Circuit simulation**

The waveforms illustrated in Figure 15 indicate the time evolution of the membrane potential of the LIF neuron, thus portraying the gradual integration of input signals and the subsequent firing of a spike. The membrane voltage is affected by the pre-synaptic input spikes in terms of both their timing and strength, as well as the nature of the synaptic connection, i.e., synaptic weight and time constants of the dynamic synapse.

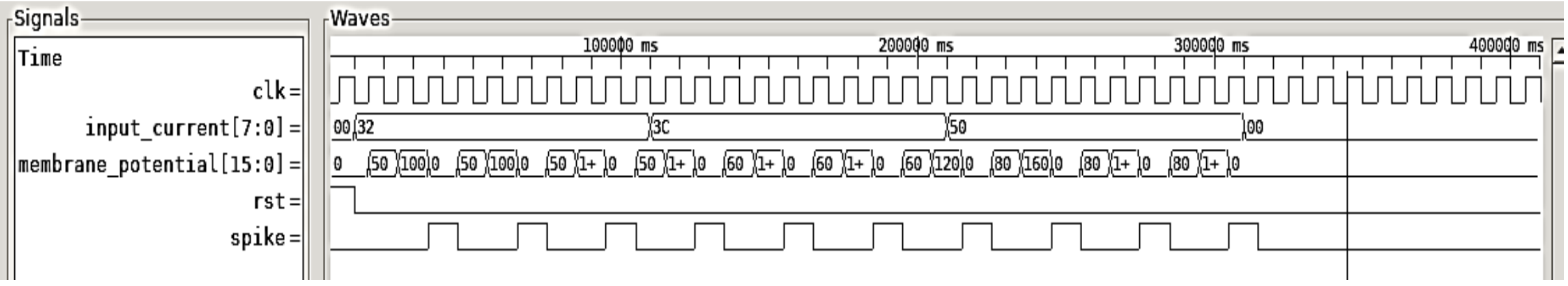

**Figure 15: Simulation of a simple spiking neuron capturing its dynamics**

### 4.2. IO Pad Integrity and Power Efficiency

In the process of developing neuromorphic VLSI chips, it is critical to take care of the I/O pads' integrity to ensure the chip's performance, especially when it is part of a larger and more complex system. These I/O pads are among the main interfaces of the chip and its external environment. The IO pads should transfer signals while the power and reliability of the signal transmission are still intact. This problem is worsened in the case of the energy-efficient neuromorphic chips, where the purpose is to get the best of both worlds. The design of the I/O pad ring is the one fundamental factor, which determines how much the chip will work efficiently overall. Our RTL to GDSII design flow in this chapter section first created the chip layout for a simple SNN LIF model and later moved to the I/O pad ring, which not only ensures minimum signal loss but also power efficiency. The I/O pad designs are discussed from different angles and ESD protection and buffer circuits are explained that control the flow of signals into and out of the chip. These circuits are essential for voltage level maintenance, without which data transfer would be inaccurate and signal degradation would result. Another very important factor in the I/O pad design is the reduction of signal reflections and noise mitigation, since both can destroy the signals' integrity and consequently, the overall performance of the chip. We therefore considered various I/O pad layouts in our neuromorphic chip simulation to mitigate this. This way we could monitor their effects on signal quality and power use.

The research indicated that a properly optimised I/O pad ring could ensure reliable signal transmission and at the same time, greatly reduce the power dissipation by cutting down on the unnecessary signal amplification. All in all, the meticulous design of the I/O pads and their related circuits guarantees that the neuromorphic chip maintains high efficiency, providing accurate and fast communication with low energy consumption, which, in turn, are significant characteristics for attaining the performance targets of contemporary neuromorphic systems.

### 4.3 VLSI Circuit Simulation Results

The simulation of neuromorphic circuits was crucial in determining both the limitations and benefits of using hardware spiking neurons. Primarily, we used Verilog code to simulate leaky integrate-and-fire (LIF) neurons with their synaptic interactions, enabling us to depict very fine waveforms corresponding to membrane potential, postsynaptic response, and output spikes, as illustrated in Figures 14 and 15. These waveforms were both crucial for confirming the dynamic synapse model's operation and providing an understanding of how the system would work in practice. The simulations also resulted in comparing different synaptic models, pointing out the most suitable way to simulate synaptic plasticity in a hardware scenario. This interpretation is very important for the tuning of the neuromorphic chips' performance, since it encourages adjusting the critical parameters, like synaptic time constants, learning rates, and the timing of spike events. Moreover, the refining of these parameters based on simulation data assures the chip not only to be the most efficient but also to replicate the biological neural networks' dynamics accurately. We then used the simulation results to analyse and see if they aligned with the expected behaviour of the LIF neuron model. The neuron membrane added up the incoming input spikes, and as soon as the membrane potential went beyond the threshold, an output spike was produced. The circuit was able to successfully emulate the fundamental neural processes of spike generation, synaptic plasticity, and temporal integration, all through the use of the same techniques that afforded enough computational efficiency and biological plausibility. A GDSII file was produced in conjunction with the Sky130nm PDK after the simulations had undergone successful verification, and the layout with pin configurations is displayed in Figure 16. The layout of the chip has its IO pins on the East, West, North, and South sides, and these pins are necessary for testing the chip's functionality: the clk, membrane potential, input spikes, spike output, and reset signals.

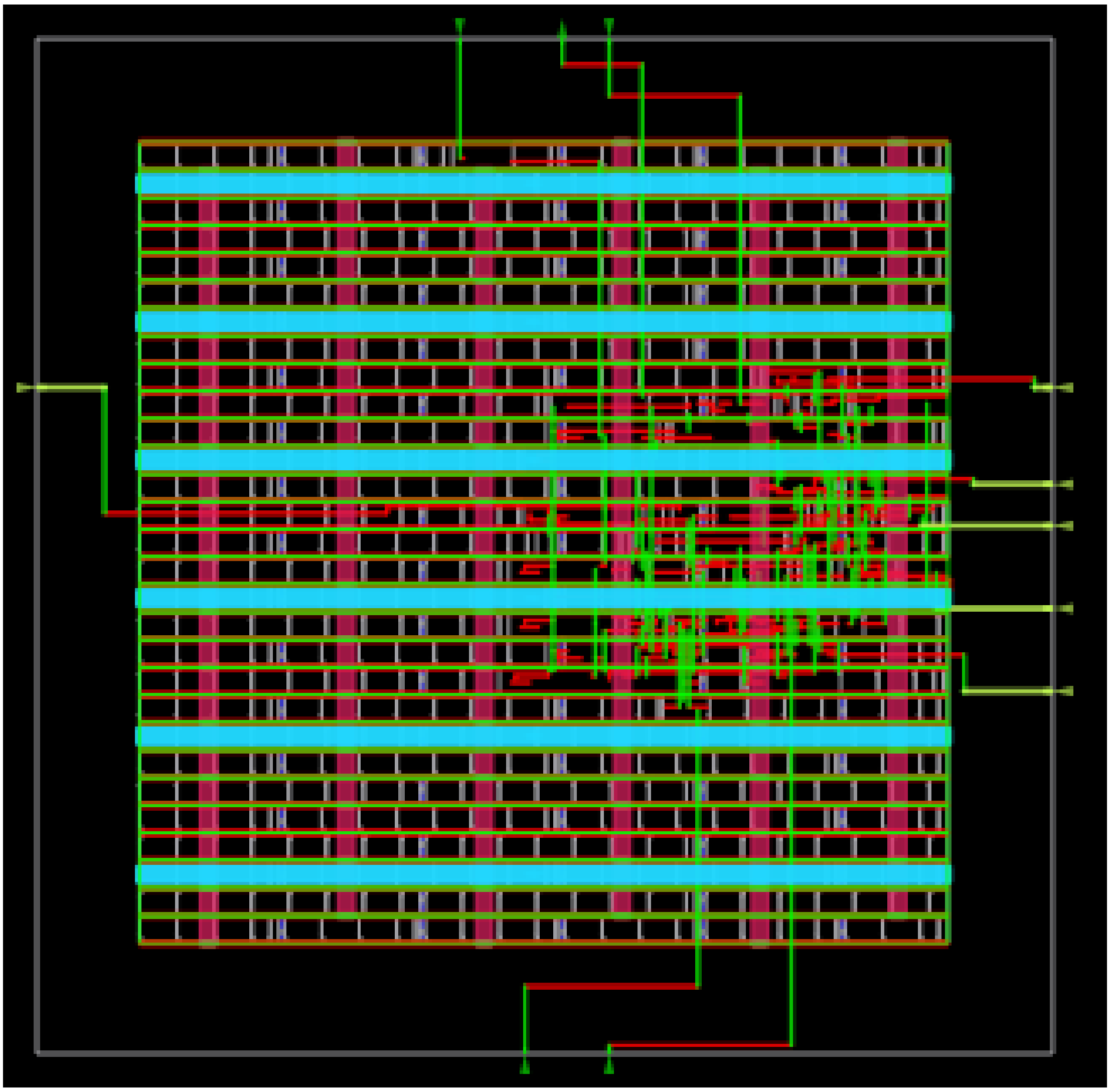

**Figure 16: GDSII Core**

It is very important to thoroughly evaluate the design synthesis statistics before adding the IO padring to the layout file. The next step contains the statistics from the design synthesis report.

**Table 3: Design synthesis report**

| Total Area | 100μm×100μm |
| --- | --- |
| cellarea (um^2) | 935.898 |
| totalarea (um^2) | 10000.000 |
| Stdcellarea (um^2) | 935.898 |
| Utilisation (%) | 14.909 |
| Peakpower (mw) | 0.120 |
| Leakage power (mw) | 0.000 |
| irdrop | 0.120 |
| Fmax (Hz) | 117.839M |
| Pins | 13 |
| Cells | 152 |
| Registers | 17 |
| Buffers | 3 |

In the context of neuromorphic VLSI design, the execution of complex models such as the LIF spiking neuron in Verilog necessitates the evaluation of the design through RTL (Register-Transfer Level) and CMOS-level synthesis reports as a step for assuring functional correctness and performance optimisation. Besides, methods like Cadence, Synopsys, and Yosys are very helpful in generating synthesis reports that could be the measure by which the design's efficiency at both logic and physical levels is weighed. The reports encompass a wide range of metrics like gate counts, timing delays, power usage, and area occupancy, making it possible to see clearly to what extent the design has complied with its intended specifications. In the process of going through the RTL synthesis reports, one is able to tell that the Verilog code is being converted into the exact number of optimised logic gates, thus maintaining the essence of the LIF spiking neuron functionality. Also, the CMOS-level synthesis reports that are generated from Yosys pinpoint the areas where performance is lagging the most. This could be in terms of timing violations or more power being consumed than necessary, both of which would influence the overall efficiency of the neuromorphic chip. Since the GDSII layout has been produced and the design statistics have been looked into, these synthesis reports serve to support the next steps in the design process, such as the crucial one of pad ring generation, which becomes the next task. This phase signifies the switch from verification and optimisation to physical implementation, which ensures that the chip is ready for the final tape-out and can operate reliably under real-world conditions. To support the need for final confirmation of the synthesis results in this phase of the design, it is very important because these results pinpoint the last changes that should be made in the system to achieve both computational accuracy and hardware efficiency in the neuromorphic system. Both RTL and CMOS level stats are provided in Tables 4 and 5 below, respectively.

**Table 4: RTL synthesis report**

| | |
|---|---|
| Number of ports: | 4 |
| Number of port bits: | 11 |
| Number of memories: | 0 |
| Number of memory bits: | 0 |
| Number of processes: | 0 |
| Number of cells: | 138 |
| $_ANDNOT_ | 51 |
| $_AND_ | 4 |
| $_DFF_PP0_ | 17 |
| $_NAND_ | 11 |
| $_NOR_ | 9 |
| $_NOT_ | 3 |
| $_ORNOT_ | 9 |

**Table 5: CMOS synthesis report**

| | |
|---|---|
| Number of wires: | **217** |
| Number of wire bits: | **254** |
| Number of public wires: | **5** |
| Number of public wire bits: | **27** |
| Number of ports: | **4** |
| Number of port bits: | **11** |
| Number of memories: | **0** |
| Number of memory bits: | **0** |
| Number of processes: | **0** |
| Number of cells: | **74** |
| sky130_fd_sc_hd__a21oi_1 | **5** |
| sky130_fd_sc_hd__a31oi_1 | **1** |
| sky130_fd_sc_hd__clkinv_1 | **18** |
| sky130_fd_sc_hd__dfrtp_1 | **17** |
| sky130_fd_sc_hd__maj3_1 | **5** |
| sky130_fd_sc_hd__nand2_1 | **4** |
| sky130_fd_sc_hd__nand4_1 | **1** |
| sky130_fd_sc_hd__nor2_1 | **4** |
| sky130_fd_sc_hd__nor3_1 | **1** |
| sky130_fd_sc_hd__nor3b_1 | **2** |

The RTL synthesis report depicted in Table 4 conveys features of the LIF spiking neuron design at the register-transfer level. The design incorporates four ports with an aggregate of 11 port bits, which corresponds to the anticipated interface specifications for the inputs and outputs of the neuron. Since there are no memories involved, it indicates that the design relies on combinational logic and registers without embedded RAM blocks. The total cell count of 138 consists mainly of basic logic gates such as AND, NAND, NOR, and flip-flops. The ANDNOT gates suggest that the synthesis tool has optimised the logic by using complex gate primitives to reduce the overall area or power consumption.

For the CMOS-level synthesis report presented in Table 5, it can be observed that the technology mapping was performed using the sky130_fd_sc_hd standard cell library. Here, the number of wires and wire bits increases, reflecting the additional routing and interconnect resources needed at the physical level. The total cell count decreases to 74, which suggests that the synthesis tool efficiently mapped the RTL logic into more complex standard cells such as AOI (AND-OR-INVERT) gates, clock inverters, flip-flops, and majority gates. This reduction in cell count was expected as physical synthesis aims to optimise the design by merging gates and minimising the number of discrete logic elements.

The uniformity in the nonexistence of memories in both synthesis reports verifies that the design consists mainly of combinational with sequential parts, but without any embedded storage blocks. The existence of clock-related components in the CMOS synthesis, such as clock inverters and flip-flops, underlines the synchronous aspect of the design, which is very important for the accurate timing of spikes and synaptic updates in the neuromorphic system. This comprehensive assessment is a prerequisite for progressing to pad ring production, as it guarantees that the design is in accordance with the performance, area, and power requirements that are crucial for the effective deployment of neuromorphic hardware.

The LIF spiking neuron model synthesis development comprised several essential stages, including the production of the activity diagrams that showed the design visually at different phases. The RTL Finite State Machine (FSM), for instance, offered a basic view of the neuron's logic and control flow, as depicted in Figure 17. It indicates which states are activated, how the transitions occur, thereby facilitating the verification of register-transfer-level correctness of the neuron's behaviour. The coarse-grained dataflow diagram provides a more abstract view of the design. This kind of diagram is particularly useful in spotting potential performance or efficiency improvement areas through its highlighting of the slower process or bottleneck. As the design approaches the actual physical implementation, the CMOS gate-level schematic is developed, which demonstrates the design corresponding to standard cells. This diagram becomes necessary since it is a representation of how the design will be created on silicon, thus giving an overview of the strength of the logic implementation at the transistor level. It ensures that the design is physically working, not only logically, and can pinpoint possible difficulties with area, power, and timing.

These diagrams and reports represent important milestones in the design process. They allow for the verification of the design functionality at every stage, from logic to physical implementation and also indicate the parts that need to be improved. Furthermore, they facilitate communication and discussion among the design team, and thus, the design will be more developed before moving to the final processes, such as pad rings generation and making a physical layout. Both RTL FSM and CMOS gate-level synthesis diagrams are shown in Figures 17 and 18, respectively.

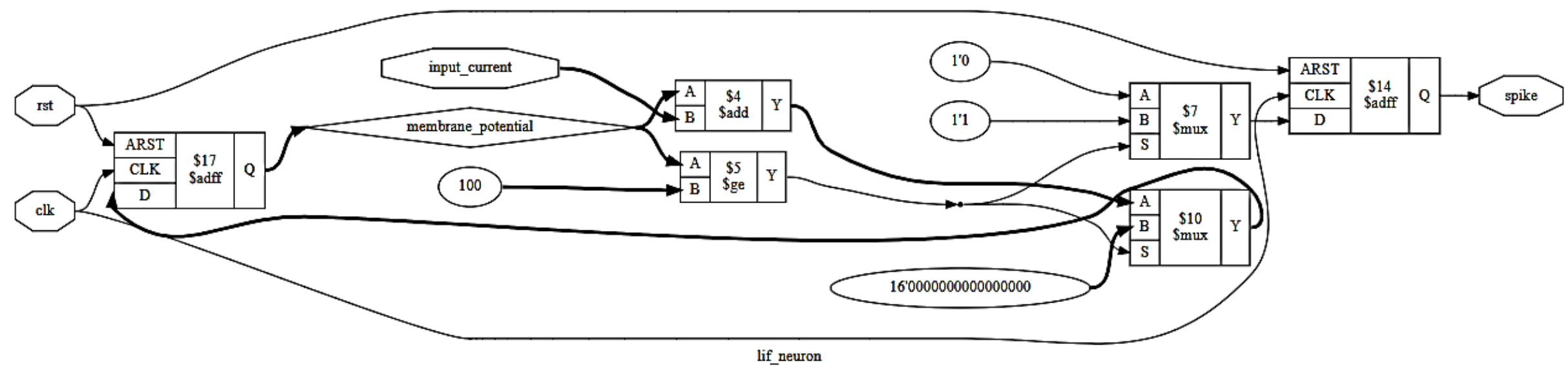

**Figure 17: RTL-level Finite State Machine and activity diagram**

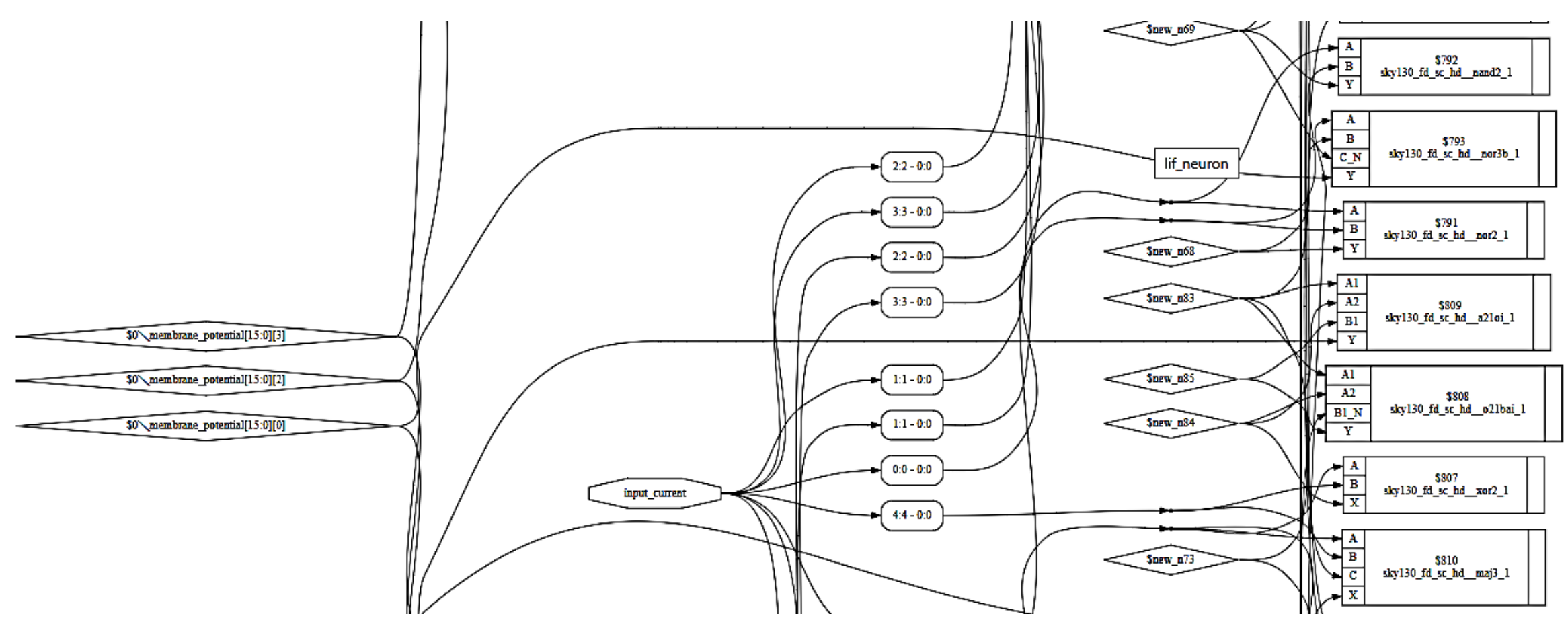

**Figure 18: CMOS snapshot activity diagram for LIF neuron model**

## 5. Input/Output Pad frame

In this section, I will elaborate on the design steps required to develop an IO pad frame that could ensure signal integrity. Once the core logic of a neuromorphic chip has been synthesised into GDSII, it may appear that the design is complete. However, without a properly engineered input/output (IO) pad frame, the circuit is not manufacturable or operable in real silicon. The pad frame defines the interface between the on-chip neuromorphic architecture and the external world, ensuring that signals, power, and ground can be delivered reliably and robustly. It provides a physical structure around the synthesised core, integrating IO pads, corner pads, and dedicated pads for power and ground.

In energy-efficient neuromorphic systems, this becomes especially critical because spiking neural networks (SNNs) require high fan-in/fan-out connectivity as well as sensitive analogue front-ends. A poorly designed pad frame can introduce signal integrity issues, ESD vulnerabilities, or power instability, all of which would hamper the architectural advantages achieved at the algorithmic and circuit level. Thus, the pad frame is a fundamental layer of physical design that must be included alongside the synthesised GDSII core to make the design functional.

### 5.1. Electrostatic Discharge Considerations (ESD)

One of the primary roles of IO pads is to provide ESD protection. Any exposed pin on a chip is vulnerable to electrostatic events that can instantaneously deliver voltages of several kilovolts. The pad frame integrates protection devices that include diodes and clamps that redirect excessive current safely to power or ground rails. A simple model for ESD protection can be written as:

$$I_{ESD} = {V_{event}}/{R_{clamp}} \qquad \text{-Eq (15)}$$

Where $V_{event}$ is the applied electrostatic voltage and $R_{clamp}$ is the equivalent resistance of the protection path. The design consideration is to keep $R_{clamp}$ low enough to shunt large transient currents without impacting normal IO behaviour. In neuromorphic systems where synaptic and neuronal circuits rely on small-signal precision, robust ESD clamps prevent catastrophic failure during packaging and testing. For this specific design of LIF neurons, a padring was developed while taking into consideration the following aspects:

- A top-level padframe that connects external chip pins to internal signals through instantiated IO pad cells. This frame must also incorporate dedicated power pads (VDD and VSS) as well as corner pads to ensure electrical and mechanical closure.
- A chip-level IO specification that defines the spatial sequence and physical arrangement of the pad cells around the die periphery.

An overview of such a padframe is shown in Figure 19.

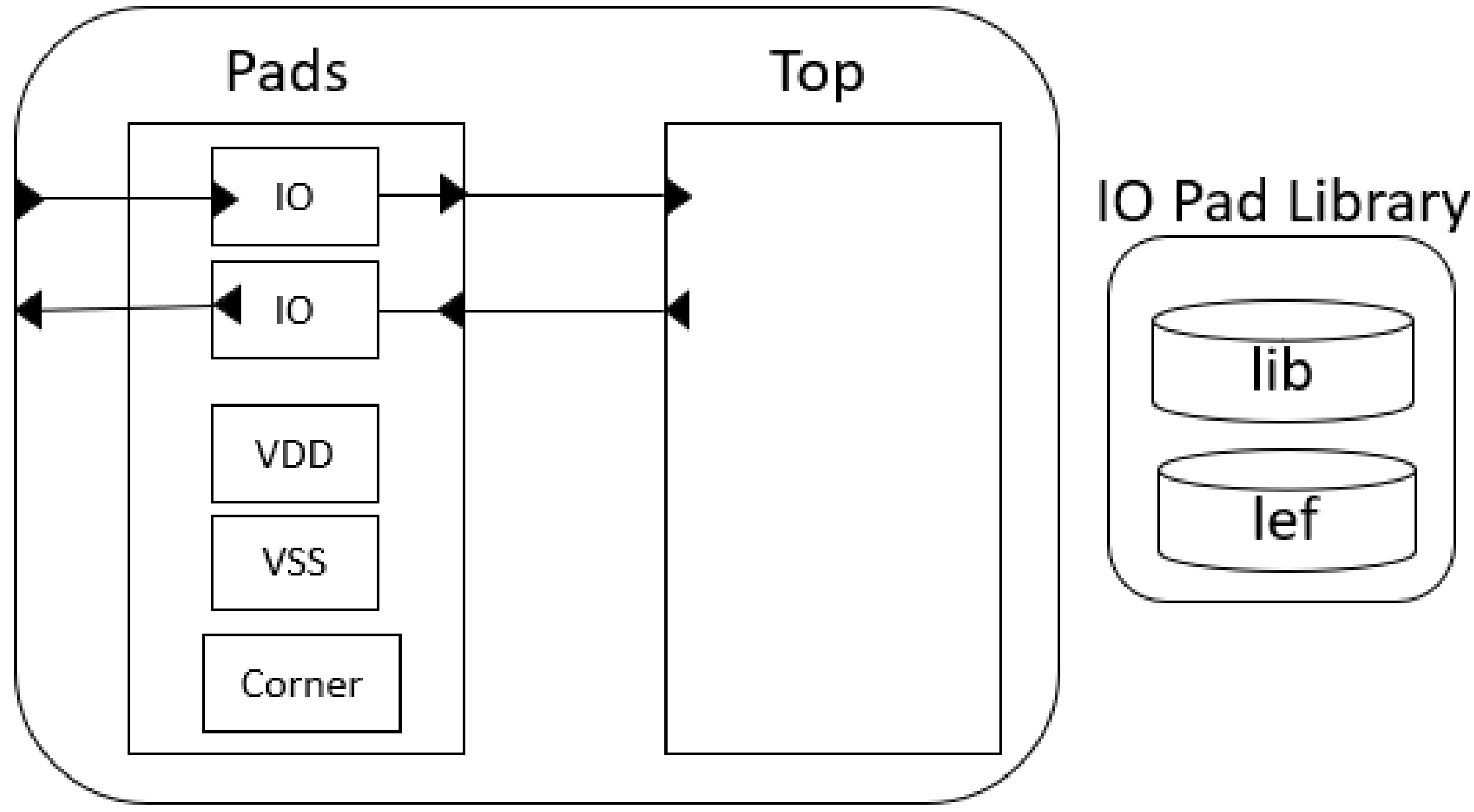

**Figure 19: Top-level design interface with Power, GND and IO pads specific to the IO Pad library provided by the PDK**

Before developing the padring, several designs were analysed with dummy pads to visualise the padframe as shown in Figure 20.

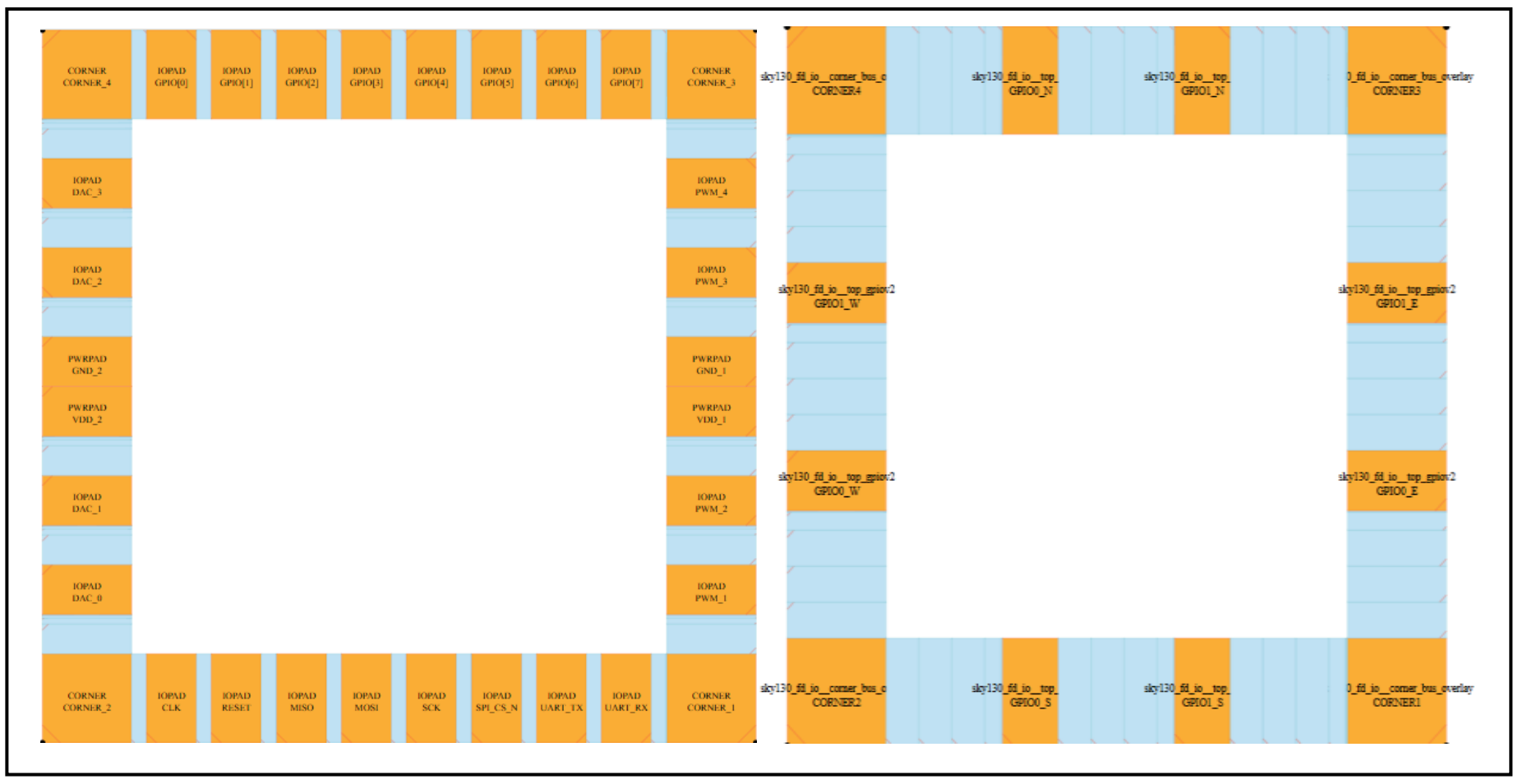

**Figure 20: Different design evaluations with dummy padrings**

Modern open-source silicon flows allow semi-automatic generation of dummy IO padrings, including corner cells and specialised pads. These padrings are crucial for ensuring mechanical closure of the chip boundary and establishing consistent power/ground delivery around the periphery. The generated padring enforces electrical symmetry across the four chip edges and provides a framework for inserting additional analogue or mixed-signal pads. Students and researchers can generate automated test padrings with tools such as Cadence and Synopsys; however, in production, these are carefully optimised for impedance, pad pitch, and package compatibility. Power integrity is a second critical function of the pad frame. The neuromorphic core may have carefully engineered clock and event-driven switching activity, but without a low-impedance supply, the activity will collapse into unstable operation. The necessary wire width for power delivery can be calculated using Ohm's law as $\Delta V = I_{load} \times R_{wire}$, where $\Delta V$ is the voltage drop along the supply rail, $I_{load}$ is the peak current drawn by the neuromorphic core, and $R_{wire}$ is the resistance of the supply trace. To maintain voltage stability, it is essential to keep this voltage drop within 5–10% of the nominal supply voltage $V_{DD}$. This imposes the constraint: $\Delta V \leq \alpha \times V_{DD}$, where $\alpha$ is a value between 0.05 and 0.10. Substituting Ohm's law into this inequality yields: $I_{load} \times R_{wire} \leq \alpha \times V_{DD}$, which can be rearranged to determine the maximum allowable wire resistance: $R_{wire} \leq (\alpha \times V_{DD}) / I_{load}$. The resistance of the wire can also be determined from its physical dimensions by the formula: $R_{wire} = \rho \times (L / (W \times t))$, where $\rho$ is the resistivity of the interconnect metal, L is the length of the power delivery path, W is the wire width, and t is the metal thickness. These formulas elucidate the role of wire shape in the resistance and voltage drop, which, hence, calls for the implementation of power rings in the chip's padring. These rings act as a barrier to the supply lines and prevent the area around them from experiencing voltage drops. When it comes to chip design, one can say that there are two major methods to create the connection between the chip and outside components. The most common one is wire bonding, which is the linking of the contact pads on the chip with the ones on the package via thin metal wires. With this method, the pads are placed at the die's outer edges. Another method, which is the flip-chip bonding, is a more modern one and it can be even higher in interconnect density. In this approach, the die is mounted face-down onto the package substrate using an array of small conductive bumps. This enables the placement of connection pads across the entire surface of the die rather than limiting them to the periphery. Figure 21 shows a typical padframe configured for both wire bonding (b) and flip-chip (a). The central region of the chip contains the core logic along with a power distribution mesh, which is enclosed by a ring of input-output pad cells and corner cells. The main function of the pad cells is to amplify low-current signals generated on the chip so they can be transmitted effectively to the outside world. Each pad cell also provides a physical location for attaching a wire in wire-bonded designs, or a solder bump in flip-chip configurations.

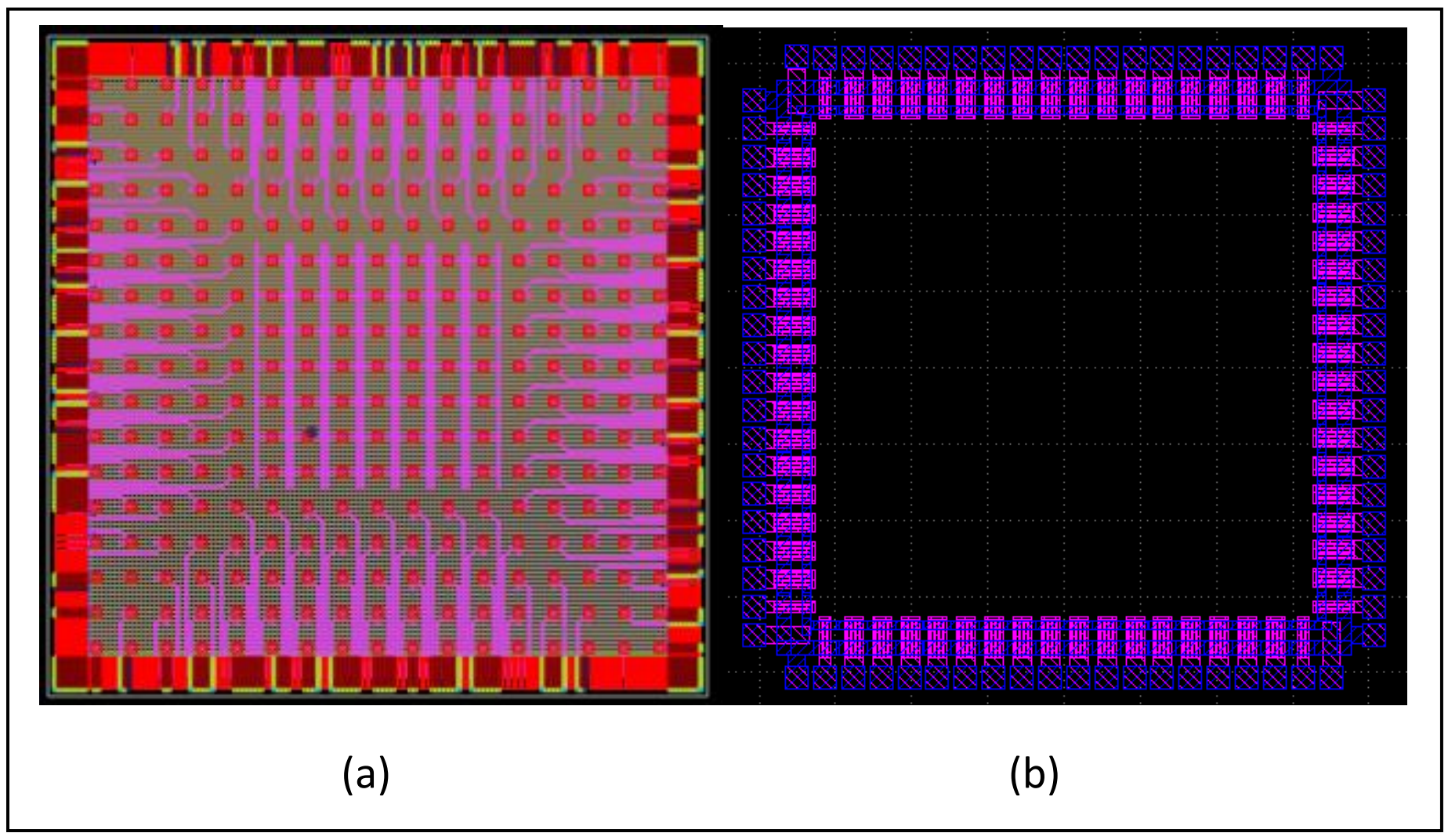

**Figure 21: An overview of the IO pad frame for core integration (a) flipchip (b)wire bonding**

In addition to signal amplification, the pad cells perform voltage level shifting when the voltage levels used for external inputs and outputs differ from the internal core voltage and those level shifters have to be integrated with the IO padframe. They also include electrostatic discharge protection, typically implemented using reverse-biased diodes between the pad and both the ground (VSS) and power (VDD) rails, to shield the internal circuitry from voltage surges [29-30]. Corner cells are included at the intersections of the pad ring to ensure that the power distribution network is continuous and properly closed. Multiple power pads, both VDD and VSS, are distributed around the chip [31-33]. These redundant power connections help ensure that power is evenly delivered across the die and that all regions maintain a stable supply voltage, as shown in Figure 22.

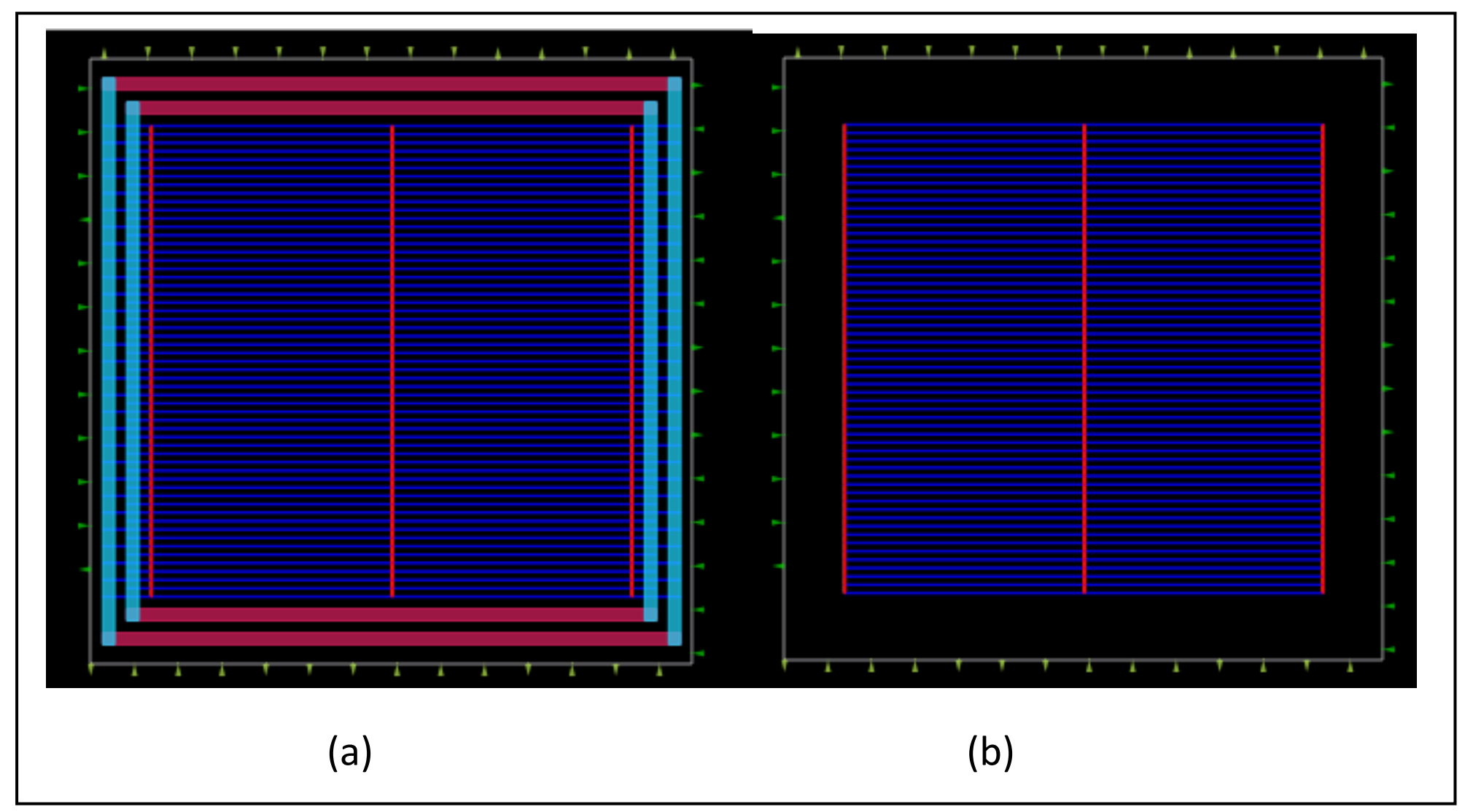

**Figure 22: (a) core with rings (b) core without rings**

## 6. Summary

The chapter discussed the important but usually overlooked part of I/O pad integrity in the context of designing energy-efficient neuromorphic chips. It is very common in the area of neuromorphic computing to concentrate on the algorithm innovations and ultra-low-power central logic, while this research places the I/O interface as a major factor in influencing the overall system performance and power efficiency. The chapter, by placing I/O pad design as a tactical factor in the development of neuromorphic VLSI systems, investigates the influence of signal reliability, power integrity, and bonding techniques on the hardware's complete functionality. Basic ideas of spiking neural networks and digital VLSI design flow are presented to give a framework. Practical knowledge is obtained from involvement with the SkyWater 130 nm CMOS processes. Key topics include I/O pad library selection, pad ring architecture, ESD protection, and bonding methods such as wire bonding and flip-chip. Through analysis and practical examples, the chapter shows how early I/O planning can significantly lower design risk, prevent costly redesigns, and foster robust neuromorphic system development. Foundational concepts of spiking neural networks and the digital VLSI design flow are introduced to contextualise the discussion. Practical implementation of the SNN model is demonstrated with the SkyWater 130 nm CMOS processes. Through theoretical analysis and practical examples, the chapter demonstrated how early-stage I/O planning can significantly reduce design risk, avoid costly redesign cycles, and support robust neuromorphic system development.

**Refe rences**

[1] Ghani, Arfan (2008), Neuro-inspired Computing Based on Area-efficient Spiking Neural Networks on Programmable Hardware, University of Ulster (United Kingdom) ProQuest Dissertations & Theses, 2008. 30524460.

[2] Gerstner W, Kempter R, van Hemmen JL and Wagner H (1996), A neural learning rule for sub-millisecond temporal coding, Nature, Vol. 383, pp. 6-78

[3] Gerstner W, Kempter R, van Hemmen JL and Wagner H (1996), A neural learning rule for sub-millisecond temporal coding, Nature, Vol. 383, pp. 6-78

[4] Gerstner W and van Hemmen JL (1994), How to describe neural activity- spikes, rates, or assemblies?, Advances in Neural Information Processing Systems 6, Morgan Kaufmann Publishers, 1994.

[5] Maass, W (1997), Fast Sigmoidal networks via spiking neurons, Neural Computation, Vol. 10, pp. 1659–1671

[6] Rieke F, Warland D, de Ruyter van Steveninck R and Bialek W, (1997), Spikes: Exploring the neural code, MIT Press.

[7] Stevens CF and Zador AM (1998), Input synchrony and the irregular firing of cortical neurons, Nature Neuroscience, Vol. 1, pp. 210–217

[8] Goodhill, G J., Perpinan, M A (2002), Cortical columns, Encyclopedia of cognitive science, McMillan publishers Ltd

[9] Izhikevich, M, E, (2004), which models to use for cortical spiking neurons, Neural Networks, Vol. 15(5), pp. 1063 – 1070.

[10] Gerstner, W., Kistler, W, (2002), Spiking Neuron Models - single neurons, populations, plasticity, Cambridge university press, UK

[11] Hodgkin, A. L and Huxley, A. F, (1952), A quantitative description of membrane current and its application to conduction and excitation in nerve, Journal of Physiology-London, Vol 117(4), pp. 500–544

[12] Gerstner, W., Kistler, W, (2002), Spiking Neuron Models - single neurons, populations, plasticity, Cambridge university press, UK

[13] Maass, W (1998), On the role of time and space in neural computation, Springer-Verlag, Vol 1450, pp. 72–83

[14] Izhikevich, M, E, (2004), which models to use for cortical spiking neurons, Neural Networks, Vol. 15(5), pp. 1063 – 1070

[15] Abbott, L F (1991), Realistic synaptic inputs for model neural networks, Network (2), pp. 245 – 258.

[16] Abbott, L F and Kepler, T B (1990), Model neurons: from Hodgkin-Huxley to Hopfield, Statistical mechanics of neural networks, Springer-Verlag, pp. 5-18.

[17] Natschläger T, Markram H and Maass W (2002), Computer models and analysis tools for neural microcircuits, A Practical Guide to Neuroscience Databases and Associated Tools, chapter 9, Kluver Academic Publishers (Boston).

[18] Maass, W, Natschläger, T and Markram, H (2002), Real-time computing without stable states: A new framework for neural computation based on perturbations, Neural Computation, Vol 14(11), pp. 2531-2560

[19] Jaeger, H. (2001), The "echo state" approach to analysing and training recurrent neural networks. Tech. Rep. Fraunhofer Institute for Autonomous Intelligent Systems: German National Research Center for Information Technology (GMD Report 148).

[20] Maass, W., Natschläger, T and Markram, H (2004), On the computational power of circuits of spiking neurons, Computer and system science, Vol 69(4), pp. 593 – 616

[21] Skowronski, MD and Harris, JG (2007), Automatic speech recognition using a predictive echo state network classifier. Neural Networks, Vol 20(3), pp. 414 – 423

[22] Hopfield, JJ and Brody, CD (2000), What is a moment? Transient synchrony as a collective mechanism for spatiotemporal integration, PNAS, Vol 98(3), pp.1282-1287.

[23] Graves, A., Beringer, N and Schmidhuber, J (2004), A comparison between spiking and differentiable recurrent neural networks on spoken digit recognition, Springer – Verlag, Vol. 31(41), pp. 127-136.

[24] Ghani, A., McGinnity, M., Maguire, L., & Harkin, J. (2008). Neuro-inspired Speech Recognition with Recurrent Spiking Neurons. In International Conference on Artificial Neural Networks (ICANN)(pp. 513-522)

[25] Izhikevich, E.M. (2003), Simple model of spiking neurons, Neural Networks, Vol 14, pp. 1569–1572.

[26] Bi G, Poo M. Synaptic modification by correlated activity: Hebb's postulate revisited. Annu Rev Neurosci. 2001;24:139-66. doi: 10.1146/annurev.neuro.24.1.139. PMID: 11283308.

[27] Ghani, A. A Proof-of-Concept Open-Source Platform for Neural Signal Modulation and Its Applications in IoT and Cyber-Physical Systems. *IoT* 2024, *5*, 692-710. https://doi.org/10.3390/iot5040031

[28] Ghani, A.; Aina, A.; Hwang See, C. An Optimised CNN Hardware Accelerator Applicable to IoT End Nodes for Disruptive Healthcare. *IoT* 2024, *5*, 901-921. https://doi.org/10.3390/iot5040041

[29] Ghani, Arfan; Dowrick, Thomas; McDaid, Liam J; (2023) OSPEN: an open source platform for emulating neuromorphic hardware. International Journal of Reconfigurable and Embedded Systems (IJRES) , 12 (1) 10.11591/ijres.v12.i1.pp1-8

[30] Morita, A. K., Takiguti, R., & Van Noije, W. A. M. (2015). *Metadata based padring and pad multiplexing generation for microcontroller design*. Journal of Integrated Circuits and Systems, 10(3), 139–146. DOI: 10.29292/jics.v10i3.416.

[31] Wang, A. Z. H., Feng, H. G., Gong, K., Zhan, R. Y., & Stine, J. (2001). *On-chip ESD protection design for integrated circuits: an overview for IC designers*. Microelectronics Journal, 32(9), 733–747. DOI: 10.1016/S0026-2692(01)00060-X.

[32] Pan, Z., Li, X., Hao, W., Miao, R., et al. (2023). *On-chip ESD protection design methodologies by CAD simulation*. ACM Transactions on Design Automation of Electronic Systems (TODAES). DOI: 10.1145/3593808.

[33] Fath, P., Moser, M., Zachl, G., Pretl, H., et al. (2024). *Open-source design of integrated circuits — an open-sourced 12-bit non-binary SAR-ADC using SKY130 PDK*. e+i Elektrotechnik und Informationstechnik (2024). DOI: 10.1007/s00502-023-01195-5.